\documentclass[structabstract]{aa}  
\usepackage[varg]{txfonts}

\usepackage{rotating}
\usepackage{graphicx}
\usepackage{tikz}
\usepackage[urlcolor=blue,citecolor=blue,colorlinks=true]{hyperref}
\usepackage[colorinlistoftodos,prependcaption]{todonotes}
\usepackage[capitalise]{cleveref}
\usepackage{booktabs}
\usepackage{lscape}
\usepackage{enumerate}
\usepackage{color}
\usepackage{amsmath}
\usepackage{siunitx}
\usepackage{xfrac}

\usepackage{longtable}
\usepackage{threeparttable} 
\usepackage{booktabs} 
\usepackage{multicol}
\usepackage{multirow}
\usepackage{xcolor}
\usepackage{ulem}

\usepackage{MnSymbol}
\def\ero{{eROSITA}}
\def\efeds{{eFEDS}}
\def\galex{{GALEX}}
\def\sdss{{SDSS}}
\def\rosat{{ROSAT}}
\def\xmm{{XMM-Newton}}
\def\chandra{{Chandra}}
\def\ixpe{{IXPE}}
\def\fuv{$FUV$}
\def\nuv{$NUV$}
\def\u{$u$}
\def\g{$g$}

\def\xstack{\texttt{Xstack}}
\def\xspec{\texttt{XSPEC}}

\def\fakeit{\texttt{fakeit}}

\def\agnsed{\texttt{AGNSED}}
\def\nthcomp{\texttt{nthcomp}}

\def\tbabs{\texttt{TBabs}}
\def\gauss{\texttt{gauss}}

\def\nh{\texttt{NH}}
\def\srctool{\texttt{srctool}}
\def\heasarc{{HEASARC}}

\def\hb{H$\beta$}
\def\mgii{$\text{Mg}\,\textsc{ii}$}
\def\civ{$\text{C}\,\textsc{iv}$}

\DeclareSIUnit\angstrom{\text{Å}}
\DeclareSIUnit\erg{\text{erg}}

\begin{document} 
\title{The Average Soft X-ray Spectra of \ero~Active Galactic Nuclei}
\author{
Shi-Jiang~Chen\inst{1,2,3}\thanks{\href{mailto:csj666@mail.ustc.edu.cn}{csj666@mail.ustc.edu.cn}}
\and
Johannes~Buchner\inst{1}
\and 
Teng~Liu\inst{2,3}
\and
Scott~Hagen\inst{4}
\and
Sophia~G.~H.~Waddell\inst{1,5}
\and
Kirpal~Nandra\inst{1}
\and
Mara~Salvato\inst{1,6} 
\and
Zsofi~Igo\inst{1,6}
\and
Catarina~Aydar\inst{1,6}
\and
Andrea~Merloni\inst{1}
\and
Qingling~Ni\inst{1}
\and
Jia-Lai~Kang\inst{2,3}
\and
Zhen-Yi~Cai\inst{2,3}
\and
Jun-Xian~Wang\inst{2,3}
\and
Ruancun~Li\inst{7,8}
\and
Miriam~E.~Ramos-Ceja\inst{1}
\and
Jeremy~Sanders\inst{1}
\and
Antonis~Georgakakis\inst{9}
\and
Yi~Zhang\inst{1}
}

\institute{
   Max-Planck-Institut für Extraterrestrische Physik (MPE), Giessenbachstrasse 1, 85748 Garching bei München, Germany
   \and
   Department of Astronomy, University of Science and Technology of China, Hefei 230026, People's Republic of China
   \and
   School of Astronomy and Space Science, University of Science and Technology of China, Hefei 230026, People's Republic of China
   \and 
   Centre for Extragalactic Astronomy, Department of Physics, Durham University, South Road, Durham DH1 3LE, UK
   \and
   Department of Physics and McGill Space Institute, McGill University, 3600 University Street, Montreal, QC H3A 2T8, Canada
   \and
   Exzellenzcluster ORIGINS, Boltzmannstr. 2, 85748, Garching, Germany
   \and 
   Department of Astronomy, School of Physics, Peking University, Beijing 100871, China
   \and
   Kavli Institute for Astronomy and Astrophysics, Peking University, Beijing 100871, China
   \and
   Institute for Astronomy \& Astrophysics, National Observatory of Athens, V. Paulou \& I. Metaxa, 11532 Athens, Greece
}

\date{Received xx, Accepted xx}

\abstract
{Active Galactic Nuclei (AGN) stand as extreme X-ray emitters where disk-corona interplay shapes their spectral energy distribution. The soft X-ray excess, a unique feature of AGN in the \qtyrange[range-phrase=--,range-units=single]{0.5}{2.0}{keV}, encodes critical information on the ``warm corona'' structure bridging the disk and hot corona. However, systematic evolution of this feature with fundamental accretion parameters in large AGN samples -- particularly studied through spectral stacking technique -- remains observationally unconstrained.}
{The \ero~All-Sky Survey (eRASS:5) provides an unprecedented sample to statistically map AGN spectral properties. We present a multiwavelength investigation of how the average AGN X-ray spectra evolve with accretion parameters ($\alpha_\mathrm{ox}$, $L_\mathrm{UV}$, $\lambda_\mathrm{Edd}$, $M_\mathrm{BH}$), and explore disk-corona connection by further combining stacked UV data.}
{We develop \xstack, a novel X-ray spectral stacking code which consistently stacks rest-frame Pulse Invariant (PI) spectra and associated responses, using optimized response weighting to preserve spectral shapes. With \xstack, we stack \num{17929} AGNs (``spec-z'' sample, total exposure $\sim \qty{23}{Ms}$) with similar X-ray loudness $\alpha_\mathrm{ox}$, UV luminosity $L_\mathrm{UV}$, and \num{4159} AGNs (``BH-mass'' sample, $\sim \qty{3}{Ms}$) with similar Eddington ratio $\lambda_\mathrm{Edd}$ and black hole mass $M_\mathrm{BH}$. The resulting stacked X-ray spectra are analyzed with a phenomenological model for both samples. We further fit the stacked optical-UV-Xray SED with the physical \agnsed~model, on a $3\times3$ $M_\mathrm{BH}$ -- $\lambda_\mathrm{Edd}$ grid.}
{The soft excess strength rises strongly with increasing $\alpha_\mathrm{ox}$ and $\lambda_\mathrm{Edd}$ binning (by a factor of \num{5}), while the hard X-ray spectral shape remains largely unchanged, consistent with the interpretation that soft excess is primarily driven by warm corona rather than reflection. Trends are weaker with $L_\mathrm{UV}$ binning and reversed for $M_\mathrm{BH}$ binning. The analysis of the optical to X-ray SEDs with \agnsed~reveals that the warm corona radius (in units of $R_\mathrm{g}$) generally increases with $\lambda_\mathrm{Edd}$ and decreases with $M_\mathrm{BH}$, or equivalently the disk-to-warm-corona transition consistently occurs near $\sim\qty{1e4}{K}$. The hot corona contracts with $\lambda_\mathrm{Edd}$ and the radius remains independent of $M_\mathrm{BH}$, aligning with disk evaporation predictions.}
{The soft excess is likely to be warm-corona dominated, with the disk-to-warm-corona transition potentially linked to hydrogen ionization instability at $\qty{\sim1e4}{K}$, consistent with previous work utilizing eFEDS-HSC stacked data \citep{Hagen+2024}. Our work highlights the power of spectral stacking for revealing AGN disk-corona connection.}
   

\maketitle


\section{Introduction} \label{sec:Intro}
The central engine of Active Galactic Nuclei (AGN) is commonly understood as composed of an accretion disk responsible for optical-UV emission \citep[e.g.,][]{Shakura&Sunyaev1973} and a hot corona responsible for hard X-ray continuum emission \citep[e.g.,][]{Haardt&Maraschi1991,Haardt&Maraschi1993}. However, the soft X-ray spectra of most unobscured type 1 AGNs show a clear excess beyond the extrapolation of hard X-ray continuum \citep[e.g.,][]{Pravdo+1981,Singh+1985,Arnaud+1985,George+2000,Reeves&Turner2000,Mineo+2000,Bianchi+2009a}. This so-called ``soft X-ray excess'' is unlikely to originate from the hot corona itself due to the soft and hard X-rays differing in variability \citep[e.g.,][]{Dewangan+2007,Noda+2011,Noda+2013,Lohfink+2013,Kara+2016a,Ursini+2020a,Mehdipour+2023,Zoghbi&Miller2023}. The soft excess temperatures are much higher than typical AGN disk temperatures \citep[e.g.,][]{Gierlinski&Done2004,Piconcelli+2005}, so that it cannot be attributed to the high energy tail of disk thermal spectrum. Furthermore, it is unlikely to be attributed to relativistically smeared partial ionized absorption \citep[e.g.,][]{Gierlinski&Done2004}, as the very high smearing speed required by data does not match theoretical predictions \citep[e.g.,][]{Schurch&Done2008,Schurch+2009a}. Its origin and formation mechanism remains a long-standing mystery.

A key observational question is how the soft excess evolves with various physical parameters. So far, research has mainly focused on individual bright sources. A clear correlation between soft excess strength and Eddington ratio ($\lambda_\mathrm{Edd}$) has been established in many works across literature \citep[e.g.,][]{Boissay+2016,Gliozzi&Williams2020,Waddell&Gallo2020,Ballantyne+2024,Palit+2024}, while no evidence has been found for the correlation between soft excess strength and black hole (BH) mass ($M_\mathrm{BH}$) \citep[e.g.,][]{Gliozzi&Williams2020}. Extending to UV, literature showed that sources with larger UV-to-Xray ratio ($\alpha_\mathrm{ox}$) tend to have steeper soft X-ray slope \citep[e.g.,][]{Walter&Fink1993,Liu&Qiao2010,Grupe+2010,Jin+2012}, though no correlation has been found between soft excess strength and the absolute luminosity \citep[e.g.,][]{Jin+2012}. While these studies are insightful, individual bright sources may not be representative of the whole population. The stacked spectra of a large sample can provide an averaged perspective, offering a complementary approach to understanding soft excess.

For the physical origin of the soft excess, two main models have been extensively discussed in recent years: the warm corona and the ionized reflection. The warm corona model attributes the soft excess to up-scattering of disk UV photons through Comptonization in a warm corona structure \citep[e.g.,][]{Magdziarz+1998,Mehdipour+2011a,Done+2012,Petrucci+2013,Petrucci+2018,Kubota&Done2018}. The ionized disk reflection posits that the soft excess arises from hard X-rays reflecting off the ionized accretion disk \citep[e.g.,][]{Ross&Fabian1993,Ballantyne+2001,Miniutti&Fabian2004,Ross&Fabian2005,Crummy+2005,Merloni+2006,Garcia&Kallman2010,Dovciak+2011,Bambi+2021}. A hybrid scenario where both contributions exist has been proved to be a more common case from both theory and observation \citep[e.g.,][]{Ballantyne2020,Petrucci+2020,Xiang+2022,Chen+2025a}. Nevertheless, the warm corona content seems to be dominant while reflection plays only a minor role (see \citealt{Waddell+2023} and \citealt{Ballantyne+2024} for X-ray spectral fitting evidence; \citealt{Porquet+2024a,Porquet+2024} for hard X-ray constraints; and \citealt{Chen+2025a} and subsequent works for UV-to-Xray analysis). Another merit of warm corona is that it naturally explains the ``UV downturn'' relative to the standard disc models and a power-law-tail bluer than $\sim\qty{1000}{\AA}$ \citep{Zheng+1997,Kubota&Done2018}, providing a self-consistent framework bridging the disk and hot corona. Under the warm corona assumption, it remains to explore how the disk transits into warm corona. 

The \agnsed~model \citep[e.g.,][]{Kubota&Done2018} enables the study of the warm corona through the comparison of the optical-UV-Xray spectrum. The model assumes no contribution from ionized reflection. The AGN disk is radially divided into three regimes: the standard multi-temperature outer disk, the Comptonization-dominated warm corona, and the hot corona. Using \agnsed~to fit the stacked \efeds~(eROSITA Final Equatorial-Depth Survey, \citealt{Brunner+2022a}) and HSC (Hyper Suprime-Cam, \citealt{Aihara+2018}) spectra, \citealt{Hagen+2024} recently reported an interesting systematic evolution of the warm/hot corona: starting from low Eddington ratio to high Eddington ratio at a constant BH mass, the hot corona component continually weakens (meaning that the hot corona radius decreases). The warm corona radius increases, however, this happens in lockstep with the disk changes, so that the disk temperature at the warm corona radius remains nearly unchanged. Such a change in optical-Xray SED is confirmed by \citealt{Kang+2025}, who further included NUV and FUV data.
Nevertheless, some limitations to the study remain. For example, they stack the unfolded X-ray spectra of individual sources, which is a model-dependent and fragile method, especially for the low-count X-ray spectra. Additionally, their BH mass was derived from the host galaxy stellar mass, which introduces large dispersion \citep[e.g.,][]{Haring&Rix2004,Kormendy&Ho2013,Reines&Volonteri2015,Shankar+2016,Davis+2018,Sahu+2019,Suh+2020,Ding+2020,Li+2021}.

The \ero~all-sky survey (eRASS; \citealt{Predehl+2021}) provides an excellent opportunity to study the above questions further. Its 2-year-long survey of the entire sky exceeds eFEDS in total exposure and total photon counts, thereby allowing a reliable constraint on both the hard X-ray continuum and the soft excess. Cross-matching eRASS with other existing multiwavelength survey catalogs such as spectroscopic quasar catalogs \citep{Wu&Shen2022} and the GALEX UV all-sky photometry \citep[][]{Martin+2005,Morrissey+2007} offers rich ancillary data on BH mass, accretion rate and UV luminosity, which are fundamental for providing a global picture on how the accretion system changes as a function of these parameters.

Technically, X-ray spectral stacking is needed to obtain a view on the average spectral shape, which is in fact non-trivial. The difficulties of stacking X-ray spectra, compared to stacking optical spectra, arise from the facts that X-ray has much fewer photon counts and generally cannot be approximated by a Gaussian distribution (meaning that the X-ray spectral counts and uncertainties cannot be scaled simultaneously), and X-ray has non-diagonal, complex response (meaning that the response needs to be taken into account when stacking). Historically, numerous X-ray stacking codes have been developed for various scientific purposes \citep[e.g.,][]{Sanders&Fabian2011,Bulbul+2014,Liu+2016,Liu+2017,Tanimura+2020,Tanimura+2022,Zhang+2024,Villalba2024}. While these codes have been sufficient for many applications, they face limitations: they may fail to (1) preserve X-ray spectral shape accurately, (2) account for photon redistribution effects when stacking the instrumental effective area curve, or (3) incorporate proper Galactic absorption corrections required for extragalactic sources studies in the soft X-ray band. Therefore, a comprehensive standalone pipeline code that incorporates all these features remains in deficit.

In this paper, we bring in our X-ray spectral stacking code, \xstack~(fully open-source\footnote{\url{https://github.com/AstroChensj/Xstack/}}), and apply it to study the average spectral properties of \ero~AGN. \cref{sec:Data} describes our sample selection and spectral extraction process. We summarize the key features of our X-ray spectral stacking code and phenomenological model in \cref{sec:X_method}, with more technical details provided in \cref{sec:Xstack} for interested readers. The UV stacking method and the physical \agnsed~model configuration are presented in \cref{sec:UV_method}. 

Using eRASS:5 data, we first analyze the averaged X-ray spectra under different physical binning choices, and present phenomenological fitting results in \cref{sec:Results_pheno}. By incorporating optical-UV stacked data, we then examine how the disk-warm/hot-corona varies with BH mass and Eddington ratio using \agnsed~(\cref{sec:agnsed_fit_results}). Finally, we compare our results with archival studies of individual sources and discuss implications for AGN disk-corona connection in \cref{sec:Discussion}. To convert flux to luminosity, we assume a cosmology with $H_0=67.7\ \mathrm{km}\ \mathrm{s}^{-1}\ \mathrm{Mpc}^{-1}$, $\Omega_\mathrm{m}=0.310$, and $\Omega_\Lambda=0.690$ \citep{PlanckCollaboration+2020}.

\section{Data\label{sec:Data}}
\begin{figure}
    \centering
    \includegraphics[width=0.9\linewidth]{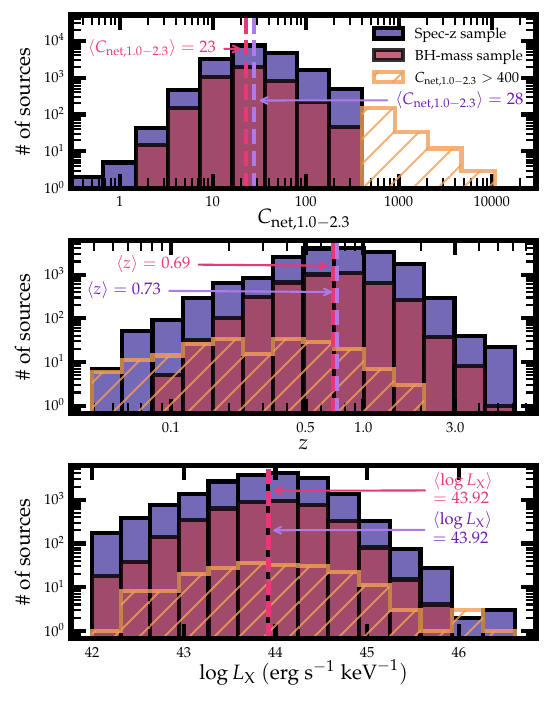}
    \caption{The distribution of \qtyrange[range-phrase=--,range-units=single]{1.0}{2.3}{keV} net counts (top panel), spectroscopic redshifts (middle panel), and \qty{2}{keV} monochromatic luminosity $L_\mathrm{X}$ (bottom panel) for our spec-z (purple filled), BH-mass (red filled), and over-bright sources in the spec-z sample (orange hatched), respectively. The median value for spec-z (BH-mass) sample is marked with purple dashed (red dashed) lines.}
    \label{fig:zcountdist}
\end{figure}

\subsection{Sample selection} \label{sec:smp_select}
Our sample is based on the first \ero~all-sky survey catalog from \citealt{Merloni+2024}. The X-ray positions are associated with optical multiwavelength counterparts (Salvato et al., in prep) based on the matching method described in \citealt{Salvato+2022}, combining NWAY\footnote{\url{https://github.com/JohannesBuchner/nway/}} \citep{Salvato+2018} together with a random forest to distinguish ambiguous cases. The optical counterpart catalog is DESI Legacy Imaging Survey \citep[][LS10 hereafter]{Dey2019}, from which we obtain the Milky-Way attenuation-corrected and deblended optical fluxes in the \g, $r$, $i$, and $z$ bands.
The counterpart catalog contains \num{761346} sources, which we further reduce with the following sample cleaning steps:
\begin{enumerate}
    \item We first select sources with \qtyrange[range-phrase = --, range-units=single]{0.2}{2.3}{keV} detection likelihood greater than \num{20} (\verb|DET_LIKE_0>20|), resulting in \num{156851} available sources. This ensures a minimum spectral quality and virtually no false detections \citep{Seppi+2022}.
    \item Then, we keep only sources with secure and best counterparts. The probability of chance association is kept low by applying 
    \verb|p_any>p_threshold8| from Salvato et al. in prep, where \verb|p_threshold8| is the threshold maximizing both completeness and purity. The purity in the relevant sky region is evaluated by placing fake sources randomly (see also \citealt{Salvato+2018}). We obtain \num{141806} sources after this step.
    \item The counterpart catalog contains not only the extragalactic sources (mostly AGN), but also galactic sources (mostly stars). Redshift information, which is essential for distinguishing between the two populations, is only available for a small fraction of sources through spectroscopic measurements, while the majority rely on photometric estimates. To construct a reliable extragalactic sample, we select sources with spectroscopic redshifts, and apply the criterion \verb|spec-z>0.002| following \citealt{Salvato+2022}, resulting in \num{32231} sources. The use of spectroscopic redshifts is also crucial for stacking X-ray spectra onto a common rest-frame wavelength grid without further losing spectral resolution due to misalignment.
    \item Next, we consider contamination by nearby sources. Due to the relatively poor spatial resolution of \ero~and \galex, some nearby bright objects could potentially be ``blended'' into the central source of interest, leading to an overestimate of the flux \citep[e.g.,][]{Bianchi+2017}. To avoid such contaminated AGN, we search for neighbors within \ang{;;6} in LS10 around our optical counterparts, where \ang{;;6} is approximately the \galex~point spread function (PSF) \citep{Morrissey+2007}. Problematic cases are UV-bright neighbors. We identify these with \verb|flux_g>3 uJy|, where we choose the \g~band flux as a proxy for \nuv/X-ray brightness. Sources below \qty{3}{\mu Jy} are much too faint to contribute with significant flux to either the optical or the UV fluxes, considering the typical optical and UV fluxes of the counterparts ($\sim\qty{90}{\mu Jy}$). Due to shredding in the LS10 catalog, mostly when a blue quasar point-like source lies on top of its extended and yellow host galaxy, and these are split into two separate LS10 sources, there are some false positives. We avoid the exclusion of these cases by keeping sources within \ang{;;0.6}, which is approximately half of the LS10 PSF size. After excluding X-ray \verb|DETUID| with counterpart information \verb|dist_arcsec<6 && flux_g>3 && dist_arcsec>0.6|, with these thresholds developed based on visual inspection, we obtain a sample of \num{28902} sources free from nearby-bright object contamination.
    \item To investigate the intrinsic soft excess of AGN, we require the AGN to be unobscured (type 1). Also, we place emphasis on the relation between the X-ray spectrum and the AGN physical properties, necessitating the use of black hole mass and bolometric luminosity. To this end, we cross-match our sample with the SDSS DR16 quasar catalog \citep{Wu&Shen2022}, which contains black hole masses estimated from \hb, \mgii~or \civ, and bolometric luminosities derived from optical SED fitting. We adopt the $M_\mathrm{BH}$ for the black hole mass if it satisfies the quality criteria recommended by \citealt{Wu&Shen2022}, preferring \hb~and otherwise relying on the \mgii~line. A total of \num{5365} type 1 AGN with a secure black hole mass measurement is obtained after this step. We note that the relatively large decrease in source number compared to step 4 is primarily because 
    only \num{8264} out of the \num{28902} sources from step 4 have been targeted by a \sdss~spectroscopic fiber (\texttt{specObj-dr16.fits}). The resulting ratio ($\gtrsim5365/8264=65\%$) indicate a 
    high type 1 fraction among the \ero~soft X-ray selected sample.
    \item The UV data offers additional clues to our understanding of the soft excess. We find \num{4481} sources lying within the \galex~survey area \citep{Bianchi+2017} and within \ang{;;2} of the optical counterpart described above, of which \num{4170} sources have \nuv~detection. The inclusion of \fuv~data is challenging, due to the extremely low detection rate and complex IGM absorption \citep[e.g.][]{Cai&Wang2023a,Cai2024}, and is therefore postponed to a future study.
    \item Finally, we drop ultra-bright sources in our sample with rest-frame \qtyrange[range-phrase = --, range-units=single]{1.0}{2.3}{keV} (the band for estimating response weights, see the top panel of \cref{sec:spec_stack}) net counts greater than \num{400} (see \cref{fig:zcountdist}), which could strongly influence the stacked spectrum and suppress the signal from the majority of dimmer sources. The resulting \num{4159} sources will be the main focus for the following sections. We analyze the bright sources individually.
\end{enumerate}
We refer to the final sample as the ``BH-mass'' sample. By construction, our ``BH-mass'' sample is constituted of only bright, broad-line type 1 AGNs. The X-ray spectra are dominated by the central AGN, and should have negligible contamination from the host galaxy (XRB) or the CGM (hot gas). The \nuv~completeness is $\sim93\%$ (compared to the \sdss~optical counterparts at step 5), and thus acceptably high. We systematically evaluate the potential selection biases from \nuv~incompleteness and redshift-dependent effects in \cref{sec:selection_bias}, and confirm that they do not significantly affect the primary conclusions in this work.


The requirement of SDSS-measured BH mass reduces our sample size sharply from $\sim$28k to $\sim$5k sources. For some dedicated analyses that do not require BH mass, we also consider a larger sample, in order to maximize the X-ray spectral quality. In our subsequent study on $\alpha_\mathrm{ox}$ and $L_\mathrm{UV}$, we define a sample that skips step 5, as we do not rely on BH mass. This, however, also removes the selection of type 1 AGN, and includes much fainter sources. To avoid heavily obscured (type 2) AGN in this sample, and to avoid low redshift galaxies with XRB dominating the X-ray emission, we additionally perform a luminosity cut in \qty{2}{keV} $L_\mathrm{X}>\qty{1e42}{erg.s^{-1}.keV^{-1}}$ and UV monochromatic luminosity $L_\mathrm{UV}>\qty{3.16e43}{erg.s^{-1}.Hz^{-1}}$ (see \cref{sec:lum_measurement} below for our derivation of UV/X-ray monochromatic flux), the boundary chosen to match our BH-mass sample. The resulting larger sample, containing \num{17929} sources, is called the ``spec-z'' sample in the sections below.

Not all type 1 AGNs are unobscured -- and some of them can even be mildly or even heavily obscured ($N_\mathrm{H}\gtrsim\qty{1e22}{cm^{-2}}$), especially for a hard X-ray selected sample \citep[e.g.,][]{Waddell+2023,Nandra+2024}. However, as can be seen in \cref{sec:Results_pheno}, our stacked spectra are typically smooth and devoid of significant absorption features ($n_\mathrm{H}<\qty{5e20}{cm^{-2}}$), suggesting that the influence from neutral absorption on our (soft X-ray selected) sample should be small in general. However, we do refer readers to \cref{sec:caveats} for a more detailed discussion on caveats and cautions regarding the absorption issues during stacking.

In \cref{fig:zcountdist} we present the distribution of rest-frame \qtyrange[range-phrase=--, range-units=single]{1.0}{2.3}{keV} net photon counts $C_\mathrm{net,1.0-2.3}$ (top), redshift $z$ (middle) , as well as the \qty{2}{keV} monochromatic luminosity $L_\mathrm{X}$ (bottom) for our sample. On average, our spec-z sample (purple histogram) has a median $C_\mathrm{net,1.0-2.3}$ of \num{28}, $z$ of \num{0.73} and $L_\mathrm{X}$ of \num{43.92}, close to our BH-mass sample (red histogram) with median $C_\mathrm{net,1.0-2.3}$ of \num{23}, $z$ of \num{0.69} and $L_\mathrm{X}$ of \num{43.92}. As a comparison, we also plot the over-bright sources (defined as $C_\mathrm{net,1.0-2.3}>400$), which are dropped in sample selection. Typically, these sources have lower redshift as compared to BH-mass or spec-z sample.

\subsection{Mass and Eddington-ratio sub-groups}
\label{sec:grid}
\begin{figure}
    \centering
    \includegraphics[width=1.0\linewidth]{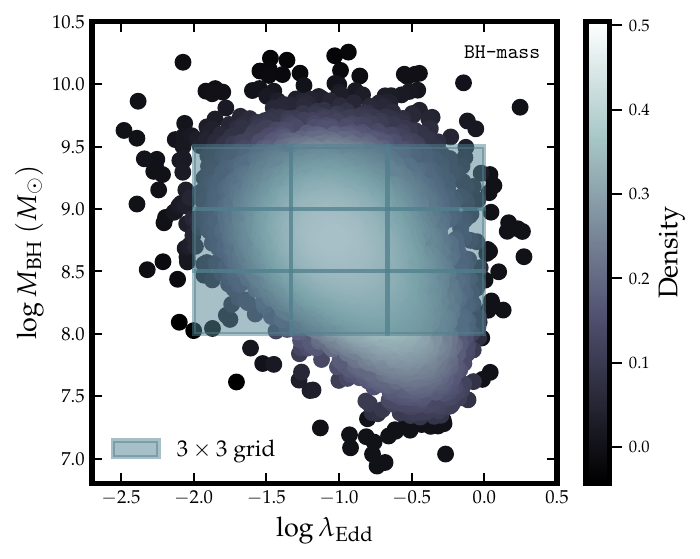}
    \caption{The distribution of $M_\mathrm{BH}$ and $\lambda_\mathrm{Edd}$ of our BH-mass sample, with color mapping density. The $3\times3$ $M_\mathrm{BH}$-$\lambda_\mathrm{Edd}$ grid is shown as blue transparent boxes.}
    \label{fig:MBHEdd_dist}
\end{figure}

The AGN optical-UV spectral shape is sensitive to BH mass and Eddington ratio, so it is crucial to stack sources with similar $M_\mathrm{BH}$ and $\lambda_\mathrm{Edd}$ when later considering stacked optical-UV data. Based on our BH-mass sample, we create a $3\times3$ grid on the $M_\mathrm{BH}-\lambda_\mathrm{Edd}$ plane, as shown in \cref{fig:MBHEdd_dist}. In each grid bin, we stack data of all sources. The $\log\lambda_\mathrm{Edd}$ binning edges are $[-2.0,-1.34,-0.67,0]$, while the $\log M_\mathrm{BH}$ binning edges are $[8.0,8.5,9.0,9.5]$. 

While stacked X-ray data should be safely AGN-dominated, this is not necessarily true for the stacked UV data. Especially the lowest $\lambda_\mathrm{Edd}$ bin with the highest $M_\mathrm{BH}$ can be contaminated by UV-luminous star formation, and we do see the stacked UV spectra are very blue (\cref{fig:agnsed_model}). With the assumption that the galaxies are on the main sequence (following \citealt{Popesso+2022}) with a galaxy stellar mass 200 times that of the black hole mass \citep{Kormendy&Ho2013}, we estimate with a simple stellar population model \cite[using GRAHSP,][]{Buchner+2024} NUV fluxes up to approximately \qty{200}{\mu Jy} for these two bins (see \cref{sec:sf_contamination} for details), which is near the median observed fluxes in those bins. While the actual contamination may be much lower if the fraction of star-forming galaxies is small, we are cautious on the interpretation of the two bins at $\log\lambda_\mathrm{Edd}<-1.5$ and $M_\mathrm{BH}>10^{8.5}M_\odot$. We do not consider these further in later discussion.

\subsection{\ero~spectra extraction}\label{sec:spec_extract}

At our sample X-ray positions (both ``BH-mass'' and ``spec-z'' sample), we extract spectra from all five \ero~all-sky surveys. This gives the deepest exposure possible, and over \num{4} times more counts than the first all-sky survey alone. The source and background spectral extraction is described in \cite{Merloni+2024}. Briefly, we feed the \ero~pipeline code \srctool{} with the calibrated event files and source coordinates. \srctool{} automatically determines the optimized source and background regions, within which the source and background spectra are extracted. Ancillary response files (ARF) are extracted from the same region as the corresponding sources, and corrected for vignetting (\verb|CORRVIGN|) and PSF loss (\verb|CORRPSF|). The response matrix files (RMF) are also extracted, but are in fact identical among all sources, since no in-orbit calibration for RMF has been taken for the moment.

\ero~consists of seven telescope modules (TMs). For TM 5 and 7, there exists a potential light leak issue due to the lack of an on-chip optical filter on the CCD \citep[e.g.,][]{Predehl+2021}, which is most pronounced below \qty{\sim 0.3}{keV} and non-negligible for soft X-ray emission line studies. For our study focusing mainly on the continuum above \num{0.3} keV, this issue should be less severe. Therefore, to improve the spectral counts we simply utilize all seven TMs. We verified that removing TM 5 and 7 did not affect our main results.

The extremely high photon counts in the \ero~stacked spectrum allow constraining the hard X-ray power-law and thereby the soft ``excess''.
The signal-to-noise ratio is high, even at the hard X-ray band. For our ``BH-mass'' sample with \num{4159} sources, we obtain total exposure time of \qty{3}{Ms} and total photon counts of \qty{3.7e5} within the rest-frame \qtyrange[range-phrase = --, range-units = single]{0.2}{10}{keV} band (\qty{7e4} within rest-frame \qtyrange[range-phrase = --, range-units = single]{2.3}{8.0}{keV} keV). For the larger ``spec-z'' sample, the total exposure time is \qty{23}{Ms} with \qty{2.1} million photon counts (\qty{4.5e5} within rest-frame \qtyrange[range-phrase = --, range-units = single]{2.3}{8.0}{keV}), likely exceeding any existing stacking experiments on AGN.

\subsection{Luminosity measurement}\label{sec:lum_measurement}
To better characterize our sample, we estimate the X-ray ($L_\mathrm{X}$) and UV ($L_\mathrm{UV}$) monochromatic luminosity for each source. Due to the complex response, the X-ray monochromatic flux (\unit{erg.s^{-1}.cm^{-2}.keV^{-1}}) or luminosity (\unit{erg.s^{-1}.keV^{-1}}) cannot be converted trivially from the Pulse Invariant (PI) spectrum (\unit{Counts.s^{-1}.keV^{-1}}), and has to be combined with spectral fitting. Therefore, to derive the rest-frame \qty{2}{keV} monochromatic luminosity $L_\mathrm{X}$ (\unit{erg.s^{-1}.keV^{-1}}), we follow a procedure similar to \citealt{Liu+2022} using \xspec~\citep{Arnaud1996}. Briefly, different models are applied to unfold the spectra, depending on the \qtyrange[range-phrase = --, range-units = single]{0.2}{2.3}{keV} net counts. If the net count is greater than 20, we adopt an absorbed power-law and blackbody model (\verb|TBabs*zTBabs*(zpowerlw+zbbody)| in \xspec~notation) on the \qtyrange[range-phrase=--,range-units=single]{0.2}{8.0}{keV} band. All parameters are optimized except for the Galactic absorption column density of \verb|TBabs|, which we fix to the $N_\mathrm{H,Gal}$ of each source\footnote{The $N_\text{H,Gal}$ values are obtained using the tool ``\nh'' provided by NASA’s High Energy Astrophysics Science Archive Research Center (\heasarc).}. When the number of counts is greater than \num{10}, we adopt the model \verb|TBabs*zpowerlaw| on the \qtyrange[range-phrase=--,range-units=single]{0.4}{2.2}{keV} band, fitting for the normalization and photon index. For the remaining cases with poor spectral quality, we fix the photon index at \num{2} and fit only for the normalization. For all cases, we read the best-fit rest-frame \qtyrange[range-phrase=--,range-units=single]{1.999}{2.001}{keV} model luminosity with \xspec~command \verb|lumin|, and convert it into \qty{2}{keV} monochromatic luminosity via $L_\mathrm{X}=\frac{L_{1.999-2.001}}{\qty{0.002}{keV}}$.

To measure the intrinsic UV monochromatic luminosity, we first perform Galactic extinction correction and K-correction on the \nuv~monochromatic flux. We assume a power-law SED ($F_\nu\sim \nu^{\alpha}$) and spectral index $\alpha=-0.65$ (\citealt{Natali+1998}; choosing a $M_\mathrm{BH}$-$\lambda_\mathrm{Edd}$-dependent $\alpha$ would not alter our results, see \cref{sec:kcorr}), applicable to AGNs within \qtyrange[range-phrase=--,range-units=single]{1200}{5100}{\AA} range:
\begin{equation}
    F_\mathrm{rest}=F_\mathrm{obs}\times 10^{0.4\times R\times E(B-V)}\times (1+z)^{\alpha-1}
\end{equation}
where we take $R=R(NUV)=7.95$ (\citealt{Bianchi+2017}, for Milky Way-type dust), and calculate $E(B-V)$ from the extinction map of \citealt{Schlegel+1998} based on the RA and DEC of each source. 
For some high-z sources (e.g., $z>1$) the adopted spectral index may not be applicable, as the rest-frame \nuv~would shift below \qty{1200}{\AA} (the lower bound of \citealt{Natali+1998}) and also IGM absorption would be stronger. Nevertheless, the impact should be minor as the high-z fraction is small. Analysing only a low-z sample also did not change our results (see \cref{sec:selection_bias} for discussion).
Then, we convert the rest-frame \nuv~monochromatic flux into rest-frame \qty{2500}{\AA} monochromatic flux, assuming the same UV SED slope. Finally, the monochromatic flux is multiplied by $4\pi d_L^2$ to derive the UV monochromatic luminosity at $2500\AA$ ($L_\mathrm{UV}$, \unit{erg.s^{-1}.Hz^{-1}}), where $d_L$ is the luminosity distance.

In order to characterize the relative importance of X-ray to UV, we adopt the ``X-ray loudness'' parameter $\alpha_\mathrm{ox}$, defined as the point-to-point slope between \qty{2}{keV} and \qty{2500}{\AA} on luminosity (\unit{erg.s^{-1}.Hz^{-1}}) diagram \citep[e.g.,][]{Avni&Tananbaum1982}:
\begin{equation}
    \alpha_\mathrm{ox}\equiv-\frac{\log L_\mathrm{UV}-\log L_\mathrm{X}}{\log\nu_\mathrm{UV}-\log\nu_\mathrm{X}}=0.3838\times(\log L_\mathrm{UV}-\log L_\mathrm{X})
\end{equation}
where the unit of $L_\mathrm{X}$ here has been converted from \unit{erg.s^{-1}.keV^{-1}} to \unit{erg.s^{-1}.Hz^{-1}} to match that of UV. Note that, following \citealt{Sobolewska+2009}, we apply an additional minus sign to keep $\alpha_\mathrm{ox}$ positive. The stronger the X-ray luminosity is compared to the UV luminosity, the smaller the value of $\alpha_\mathrm{ox}$ would be.

The luminosity measurements come with uncertainties. For our BH-mass sample (similar for spec-z sample), the median \qty{68}{\%} uncertainty in $\log L_\mathrm{X}$ is $\sim\qty{0.14}{dex}$, while the photometric uncertainty in $\log L_\mathrm{UV}$ is $\sim\qty{0.13}{dex}$. After propagating these errors, the median uncertainty in $\alpha_\mathrm{ox}$ is $\sim\num{0.075}$. Notably, these uncertainties for individual sources remain nearly half as small as the typical bin size of each subgroup ($\sim\qty{0.34}{dex}$ for each of the 3 $L_\mathrm{UV}$ subgroups, and $\sim\qty{0.14}{dex}$ for each of the 3 $\alpha_\mathrm{ox}$ subgroups). Therefore, they are unlikely to significantly impact our later conclusions involving $L_\mathrm{UV}$ or $\alpha_\mathrm{ox}$.

\section{Methods for X-ray spectral stacking and fitting}\label{sec:X_method}

\subsection{\xstack: a novel tool for X-ray spectral stacking}\label{sec:Xstack_pointer}

Given the large number yet individually low-count nature of X-ray spectra in the era of \ero, a convenient and robust tool for X-ray spectral stacking sources from different redshifts is urgently needed, to support a wide range of scientific investigations. Although X-ray spectral stacking is a classical technique for improving signal-to-noise ratio and has already been applied to many scientific problems \citep[e.g.,][]{Sanders&Fabian2011,Bulbul+2014,Liu+2016,Liu+2017,Tanimura+2020,Tanimura+2022,Zhang+2024,Villalba2024}, certain technical limitations still exist. These include, how to optimally preserve spectral shape when stacking sources with different redshifts and fluxes; how to properly account for Galactic absorption during stacking to recover the intrinsic extragalactic spectra; and how to perform statistically sound spectral fitting on the stacked result.

To address these issues, we develop \xstack, a standalone pipeline for X-ray spectral shifting and stacking. Briefly, \xstack~first sums all (rest-frame) PI spectra, without any scaling; and then sum the rest-frame corrected full responses (ARF*RMF), each with appropriate weighting factors to preserve the overall spectral shape. 

The more technical aspects of \xstack~are presented in \cref{sec:Xstack}, beginning with a pedagogical review of basic concepts in X-ray spectroscopy. Here, we highlight the key features of the code:
\begin{enumerate}
    \item \textit{Preservation of spectral slope,} achieved by properly assigning weighting factors to each response (ARF*RMF) during the stack;
    \item \textit{Correction for Galactic absorption,} by applying a $N_\mathrm{H,Gal}$-dependent Galactic absorption profile (\tbabs) to each ARF before shifting and stacking (see also \citealt{Zhang+2024});
    \item \textit{Well-defined statistics for subsequent spectral fitting,} due to the preservation of Poissonian nature of the data. The stacked background is approximated with a Gaussian distribution.
\end{enumerate}
In \cref{sec:simulation}, we also validate our code's ability to preserve spectral shapes through simple power-law simulations.

\subsection{X-ray Spectral fitting: global settings}\label{sec:spec_fit}

The correct propagation of uncertainty allows us to perform spectral fitting on the stacked PI spectrum, combined with the stacked response. 
Each stacked spectrum considered is analyzed within the \qtyrange[range-phrase=--,range-units=single]{0.3}{8.0}{keV} band. Below \qty{0.3}{keV}, we find that only $\lesssim 20\%$ sources contribute to the rest-frame stacked spectra ($\lesssim 30\%$ sources contributing to the rest-frame stacked ARF), potentially leading to an insufficient sampling of ARF and background data; while above \qty{8}{keV} we find that the background contribution is higher than $\sim 90\%$ and does not provide much information. 
Since our data follow the Poisson distribution while the scaled sum of backgrounds approximates the Gaussian distribution, we adopt the PG statistic\footnote{\url{https://heasarc.gsfc.nasa.gov/xanadu/xspec/manual/XSappendixStatistics.html}}. The parameter confidence regions are derived from the 100 bootstrapped stacked spectra fit, which accounts for both Poisson fluctuation and sample variance.

\subsection{Phenomenological spectral model}\label{sec:pheno_model}
To characterize the strength of soft excess in model-independent way and compare with literature, we first apply a phenomenological rest-frame model, which can be written as \verb|TBabs*(powerlaw+cutoffpl+gauss)| in \xspec~terminology. We now describe the components of this model in turn.

The hard X-ray primary continuum is modeled by a \verb|powerlaw|, $N=A\times E^{-\Gamma_\mathrm{PC}}$\footnote{Note, $N$ here is photon flux in units of \unit{Photons.cm^{-2}.s^{-1}.keV^{-1}}.}, with the photon index $\Gamma_\mathrm{PC}$ set within a relatively narrow range \numrange[range-phrase=--]{1.7}{2.3}, because (1) from a data-driven point of view, the hard X-ray spectral index of our stacked spectra in \cref{fig:pheno_data} do not deviate much from 2, and we want to prevent the rare case where an unphysically hard ($\Gamma_\mathrm{PC}\lesssim1.5$) hard X-ray primary continuum and an ultra-strong soft excess is derived from the fit; (2) the range \numrange[range-phrase=--]{1.7}{2.3} is reasonable compared with literature \citep[e.g.,][]{Nandra&Pounds1994,Brightman+2013a}, given the median physical quantity of each $\alpha_\mathrm{ox}/L_\mathrm{UV}/\lambda_\mathrm{Edd}/M_\mathrm{BH}$ subgroup. A Gaussian component \gauss~is also included to account for the iron line at $\sim \qty{6.4}{keV}$, but since it is not the major focus of this study, we simply fixed the line centroid at \qty{6.4}{keV} and sigma at \qty{0.019}{keV} \citep[e.g.,][]{Shu+2011}. 

A cut-off power law is adopted to model the soft excess. The spectral model is $N=A\times E^{-\Gamma_\mathrm{SE}}\exp\left(-E/E_\mathrm{cut}\right)$, with the photon index $\Gamma_\mathrm{SE}$ set to be softer than $\Gamma_\mathrm{PC}$ by at least 0.5 ($\Gamma_\mathrm{SE}-\Gamma_\mathrm{PC}>0.5$), and the cut-off energy $E_\mathrm{cut}$ set below \qty{2}{keV}. The former setting is to ensure a physically interpretable result, while the latter setting is motivated by observed soft excess spectra showing a bend-over at high energies (\cref{fig:spec_diff}). Furthermore, if the soft excess component is a ``warm corona'' structure with typical temperature within \qtyrange[range-phrase=--,range-units=single]{0.1}{1.0}{keV} \citep[e.g.,][]{Petrucci+2018}, its spectrum is approximately a cut-off power law with the cut-off energy $E_\mathrm{cut}$ by a factor of $\sfrac{1}{2}\sim\sfrac{2}{3}$ lower than the temperature. Following the standard in literature \citep[e.g.,][]{Boissay+2016}, we also measure the strength of the soft excess component, defined as the luminosity ratio of soft excess to primary continuum within the \qtyrange[range-phrase=--,range-units=single]{0.5}{2.0}{keV} band:
\begin{equation}
    q\equiv\frac{L_\mathrm{cutoffpl,0.5-2}}{L_\mathrm{powerlaw,0.5-2}}
\end{equation}

Finally, all the above components are absorbed by an intrinsic absorption layer (\tbabs). The column density $n_\mathrm{H}$ is allowed to vary between $0\sim\qty{10e20}{cm^{-2}}$, a reasonable range for optically unobscured AGN \citep[e.g.][]{Ricci+2013a}. We tested the \verb|disnht| model \citep{Locatelli+2022}, which allows a log-normal distribution of column densities, but the standard deviation parameter is difficult to constrain when a soft excess component is also modeled.

\section{Methods for UV stacking}\label{sec:UV_method}
\subsection{UV photometry stacking}\label{sec:UV_stack}
For a panchromatic understanding of the underlying physics, it is necessary to combine UV data in spectral fitting, and apply a more physical model.

To that end, we stack the \galex~\nuv, \sdss~\u~and \g~\citep[][]{Abazajian+2009} photometry for each source, after galactic extinction correction (we take $R(u)=4.24,\ R(g)=3.30$ from \citealt{Schlafly&Finkbeiner2011}) and K-correction with the same assumption as in \cref{sec:lum_measurement}. That means for a source with $z=0.5$, its \nuv~flux with central wavelength at \qty{2315.7}{\AA} will still contribute to a data point at \qty{2315.7}{\AA} in the stacked spectrum, rather than at rest-frame wavelength $2315.7/(1+0.5)=\qty{1543.8}{\AA}$. We do not stack the rest-frame flux directly, due to (1) in the energy range where both \sdss~and \galex~contribute, the \sdss~signal will completely dominate over \galex, as \sdss~has much larger effective area than \galex; (2) because we are stacking discrete UV photometry with relative narrow wavelength range compared to redshift distribution, the overlapping region for different sources on the rest-frame energy grid will be small, leading to an overall small completeness in each energy bin; (3) for high redshift sources contributing to the rest-frame EUV, their stacked spectrum is severely biased due to IGM distortion. Additionally, we note that using a $M_\mathrm{BH}$-$\lambda_\mathrm{Edd}$-dependent UV SED slope for K-correction does not significantly affect our later \agnsed~fit, with changes in the best-fit parameters remaining within $1\sigma$.

We now have many individual UV photometry points anchored on energies corresponding to \g~(\qty{2.6}{eV}), \u~(\qty{3.5}{eV}) and \nuv~(\qty{5.4}{eV}). Since the effective areas for each energy bin do not vary significantly from source to source, we sum the flux in each energy bin. To ensure that the stacked UV spectrum can be compared with stacked X-ray spectrum, we divide the entire stacked UV spectrum by the X-ray response normalization factor ($\sum_l W^l$ in \cref{sec:spec_stack}). 

Photometry often comes with calibration issues \citep[e.g.,][]{Arnouts&Ilbert2011,Bianchi+2017}, which cannot be improved by stacking. Moreover, some systematic issues (e.g., IGM absorption, and uncertainty of K-correction) degrade the reliability of our UV data. Therefore, we follow the standard practice of SED fitting procedure, by accounting for these extra uncertainties as $\eta$ fraction of stacked flux ($F_{g/u/NUV}$), and add them in quadrature in addition to the original uncertainty ($\sigma_{g/u/NUV}$) propagated from the catalog:
\begin{equation}
    \sigma_{\mathrm{Total},g/u/NUV}=\sqrt{\sigma_{g/u/NUV}^2+(\eta\times F_{g/u/NUV})^2}
\end{equation}
In practice, $\eta$ is chosen as 1.5\% for \g~and \u~band. For \nuv, we conservatively choose a value as large as 30\% \citep[e.g.,][]{Georgakakis+2017}, as the calibration and systematic issues are generally more severe in \galex. Finally, we note that decreasing the statistical weight of UV data in UV-Xray linked spectral fitting could also prevent the UV data (typically with smaller uncertainties compared with X-ray) from dominating the fit. 

At the end of each bootstrap, we also add additional random scatter to the data (not only the error bar) with amplitude corresponding to $\eta$ fraction of stacked flux, in order to account for the systematic issues.

\subsection{Physical spectral model: \agnsed}\label{sec:fit_agnsed}
We fit our optical-UV-Xray stacked data with the physical model \agnsed~\citep{Kubota&Done2018}. In this model, the accretion disk is radially separated into three regimes. For a given black hole mass and Eddington ratio, the standard Novikov-Thorne emissivity profile \citep[][]{Novikov&Thorne1973} is assumed for the accretion disk, which extends as far as $R_\mathrm{out}$. At each annulus, the emergent spectrum is in the form of blackbody. Below the radius $R_\mathrm{WC}$, the Compton scattering in the disk surface becomes dominant, and the disk transits into a ``warm corona'', which is modeled by a Comptonized component (\nthcomp) with photon index $\Gamma_\mathrm{WC}$ and hot electron temperature $kT_\mathrm{WC}$. The seed photon temperature, on the other hand, varies with radius and is tied to the local underlying disk temperature.
Going further down to $R_\mathrm{HC}$ ($R_\mathrm{HC}<R_\mathrm{WC}$), the ``warm corona'' or disk is believed to truncate, forming a ``hot corona'' with completely different physical properties, characterized by a hot Comptonized component with photon index of $\Gamma_\mathrm{HC}$ and electron temperature $kT_\mathrm{HC}$. 

Our physical spectral model is written as \verb|TBabs*(agnsed+gauss)| in \xspec~terminology. We fix the $M_\mathrm{BH}$ ($\lambda_\mathrm{Edd}$) at the sample median value of each bin. For the warm corona, we allow $\Gamma_\mathrm{WC}$ to vary between \num{2.25} and \num{5}, and $kT_\mathrm{WC}$ from \qtyrange[range-phrase=--,range-units=single]{0.1}{1.0}{keV}. For the hot corona, we fix $kT_\mathrm{HC}$ at \qty{100}{keV}, while allowing for $\Gamma_\mathrm{HC}$ to vary between \num{1.7} and \num{2.3} for the same reason as in \cref{sec:pheno_model}. Following \citealt{Hagen+2024}, we fix $R_\mathrm{out}$ at $10^4\ R_\mathrm{g}$ (gravitational radius, $R_\mathrm{g}=\frac{2GM_\mathrm{BH}}{c^2}$), BH spin $a_*$ at \num{0.0}, cosine of inclination angle at \num{0.87} (as the sources are type 1), and hot coronal height at $10\ R_\mathrm{g}$. Other parameters of \agnsed, including $R_\mathrm{WC}$ and $R_\mathrm{HC}$, are allowed to vary freely. For the intrinsic absorption \tbabs~and iron line complex \gauss, we adopt a similar setting as in \cref{sec:pheno_model}.

\agnsed~is an energy-conserving model, and for fitting individual sources (rather than the stacked spectrum), its normalization is typically fixed at \num{1}, with the distance set to the physical value corresponding to the source redshift. However, due to the nature of our stacking (which preserves spectral shape rather than absolute normalization), and also because we are primarily interested in the spectral shape, we adopt an arbitrary distance and allow the normalization to vary freely. We note that this approach is only valid, when stacking sources with similar $M_\mathrm{BH}$ and $\lambda_\mathrm{Edd}$ (and thus comparable luminosities), as is the case for our BH-mass sample in \cref{sec:agnsed_fit_results}.

\section{Phenomenological modeling of stacked X-ray spectra}\label{sec:Results_pheno}
\subsection{Full sample results}
\begin{figure}
    \centering
    \includegraphics[width=1.0\linewidth]{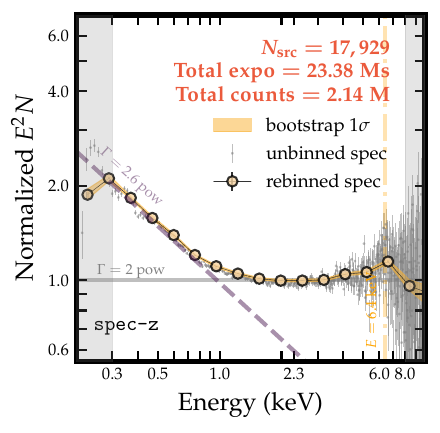}
    \caption{Summed \ero{} spectrum from \qty{23}{Ms} of observations (``spec-z'' sample). To reveal the intrinsic spectral shape of AGN, the spectrum is divided by the ARF and multiplied by $E^2$ (so the unit proportional to \unit{keV^2.Counts.cm^{-2}.s^{-1}.keV^{-1}} or \unit{erg.s^{-1}}), normalized at \qty{4}{keV}. The gray points are the unbinned data, while yellow circles show a logarithmic rebinning. The yellow shaded area, derived from bootstrapping the sample for 100 times, indicates the $68\%$ uncertainty range of the spectrum from both stacking process and Poisson fluctuation (the plotting routine also applies to \cref{fig:pheno_data,fig:spec_diff,fig:pheno_model_alphaoXLUV,fig:pheno_model_EddMBH,fig:agnsed_model}). Only the spectrum between \qty{0.3}{keV} and \qty{8}{keV} is used in this work. Between \num{1} and \qty{6}{keV}, the spectrum is approximately linear in this log-log plot, while below, the spectrum bends upwards. Between \num{6} and \qty{7}{keV}, a bump is visible despite the relatively larger errorbars. Power-laws of typical photon indices are plotted only for visual comparison.}
    \label{fig:c038_data}
\end{figure}
\begin{figure*}
    \centering
    \includegraphics[width=1.0\linewidth]{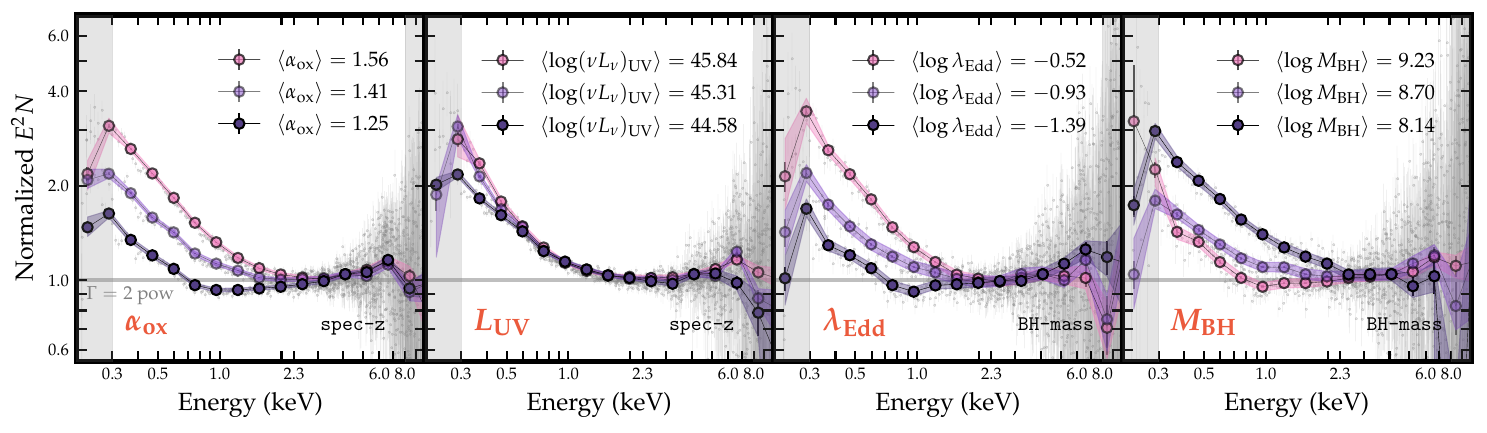}
    \caption{Stacked spectra in three equally sized groups (colored lines), divided by $\alpha_\mathrm{ox}$, $L_\mathrm{UV}$, $\lambda_\mathrm{Edd}$, $M_\mathrm{BH}$ (panels from left to right). The legend gives the mean value of each sub-group. Error bars are obtained with bootstrapping as in \cref{fig:c038_data}, but are generally very small except above $\sim \qty{8}{keV}$. In the left panel, the three colored lines are indistinguishable above \qty{2}{keV}. Below \qty{1}{keV}, the blue data points from the lowest $\alpha_\mathrm{ox}$ sub-sample are always below the pink points from the highest $\alpha_\mathrm{ox}$ sub-sample. The $\lambda_\mathrm{Edd}$ panel shows an even larger spread between sub-groups, while it is less pronounced in the $L_\mathrm{UV}$ panel and reversed in the $M_\mathrm{BH}$ panel.}
    \label{fig:pheno_data}
\end{figure*}
\begin{figure}
    \centering
    \includegraphics[width=1.0\linewidth]{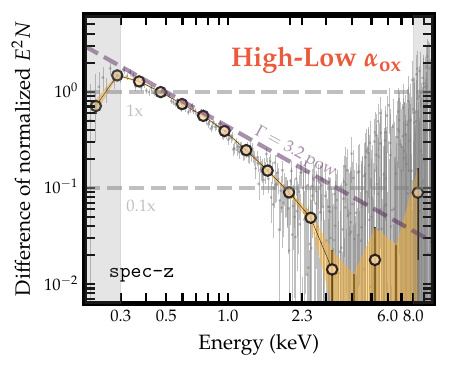}
    \caption{Difference spectrum between the highest and lowest $\alpha_\mathrm{ox}$ subgroups from the left panel of \cref{fig:pheno_data}, highlighting the spectral shape of the emerging soft excess component. The binned yellow data follow a line in this log-log plot corresponding to a $\Gamma=3.2$ power-law (dashed purple line), but fall below beyond \qty{1}{keV}. 
    The horizontal lines mark where the flux ($E^2N$) of the difference spectrum amounts to 100\%, or 10\% of the original normalized spectra at \qty{4}{keV}, where the spectra are normalized. The yellow data points are above the 1x line below \qty{0.5}{keV}, and below the 0.1x line above \qty{2}{keV}. Above \qty{3}{keV}, the difference falls below 1\%.
    }
    \label{fig:spec_diff}
\end{figure}
\begin{figure*}
    \centering
    \includegraphics[width=0.9\linewidth]{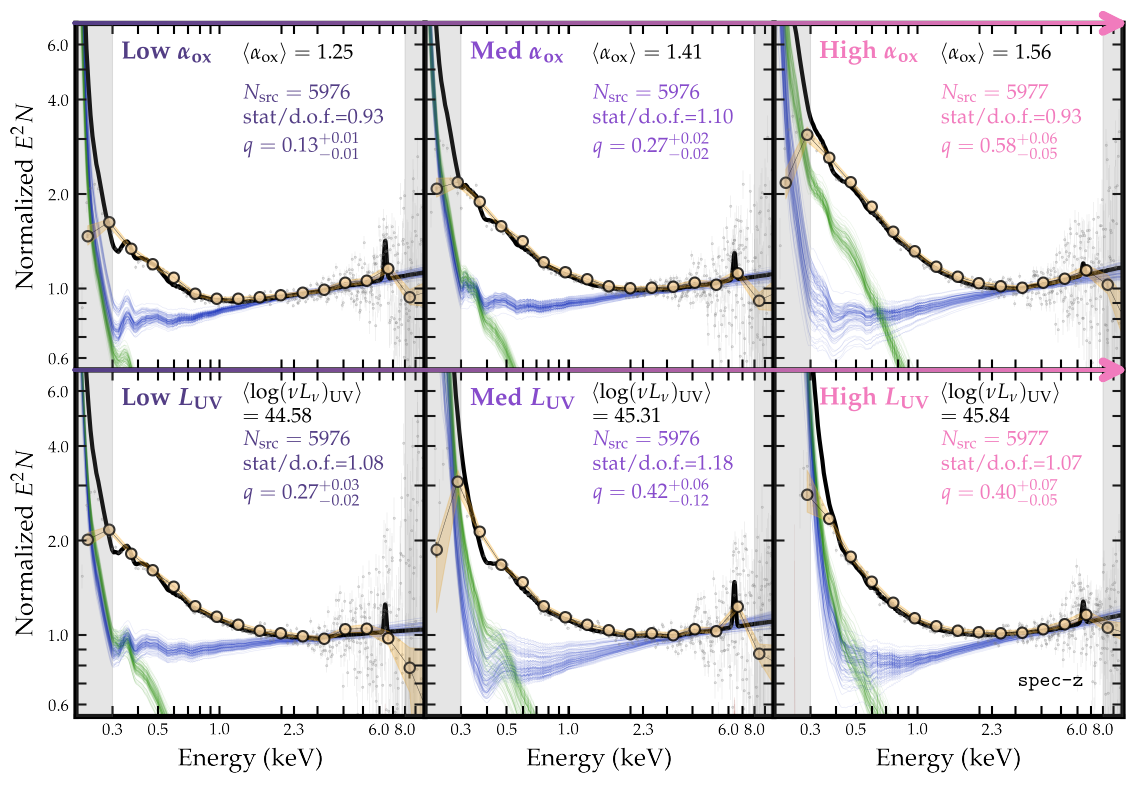}
    \caption{Fits to stacked spectra (from ``spec-z'' sample) in equally sized bins of $\alpha_\mathrm{ox}$ (top panels) and $L_\mathrm{UV}$ (bottom panels). The legend gives the mean value of each sub-group, the number of stacked sources, and the fitting statistics. The soft power-law (green) and the hard power-law (blue), as well as the sum (black) are shown. Each curve is the best fit to a bootstrap realization of the sample.
    In the top panels, from the left to right panels, the green curves move upwards. The strengthening of the soft excess is quantified by $q$ with error bars, increasing from \num{0.13} to \num{0.58}. The bottom panels show less change of the green warm corona component with $L_\mathrm{UV}$.
    }
    \label{fig:pheno_model_alphaoXLUV}
\end{figure*}
\begin{figure*}
    \centering
    \includegraphics[width=0.9\linewidth]{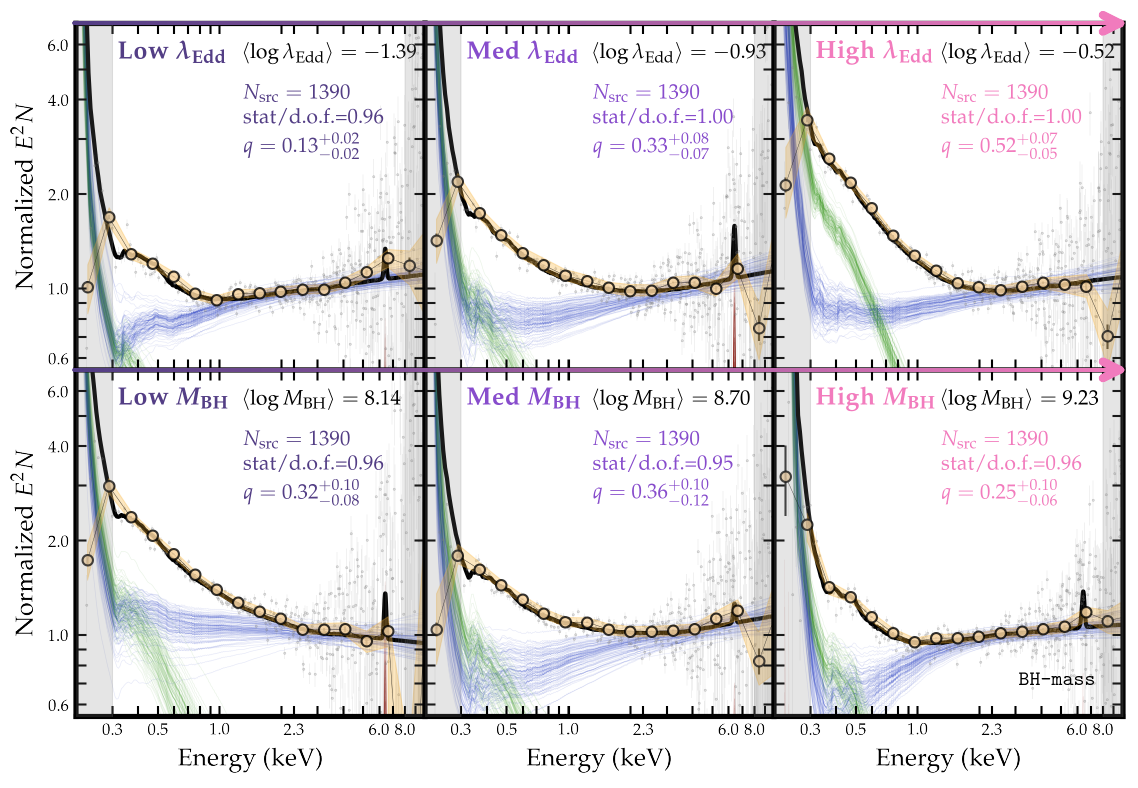}
    \caption{Like \cref{fig:pheno_model_alphaoXLUV}, but for $\lambda_\mathrm{Edd}$ (top panels) and $M_\mathrm{BH}$ (bottom panels), using the ``BH-mass'' sample. In the top panels, from the left to right panels, the green soft excess component curves move upwards, and the soft excess strength $q$ increases from \num{0.13} to \num{52}.
    }
    \label{fig:pheno_model_EddMBH}
\end{figure*}
In \cref{fig:c038_data}, the stacked spectrum of \num{17929} sources from our largest spec-z sample is presented. The spectrum is smooth and appears bent upwards towards the soft end. To reveal the intrinsic spectral shape unaffected by the varying effective area and channel energy width across the whole energy band (\qtyrange[range-phrase=--,range-units=single]{0.2}{10}{keV}), the plot shows the stacked spectrum divided by the effective area (from the stacked ARF) and energy width of each channel. Since we want to highlight the ``soft excess'' from the underlying hard X-ray power-law (with a typical photon index of $\sim \num{2}$), we multiply the spectrum by the energy squared to obtain units of \unit{keV^2.Counts.cm^{-2}.s^{-1}.keV^{-1}} ($E^2N$). Finally, as we are not interested in the absolute normalization for this visualization, we renormalize the resulting ``ARF-weighted'' spectrum at 4 keV, where the hard X-ray power-law dominates, shown as gray points in \cref{fig:c038_data} (``unbinned spec''). For clarity, we also rebin the spectrum on a logarithmic energy scale, as shown by the yellow-filled circles (``rebinned spec''). The uncertainty range of the rebinned stacked spectrum, resulting from both stacking process and Poisson fluctuation, is computed by 100 bootstraps, which randomly leave out spectra from the sample. This is shown as the yellow shaded area in \cref{fig:c038_data}. Since both the number of sources and the number of photon counts are very large, the uncertainties are very small. Because bootstrapping error bars are sensitive to the heterogeneity of the sample, finding small error bars indicates a similar spectrum within the stacked sample.

Several notable features are apparent in the binned spectrum in \cref{fig:c038_data}. Between \qty{1}{keV} and \qty{6}{keV}, a primary power-law component with a photon index of $\sim \num{2}$ is evident. This is typical for AGN and consistent with the average value observed in eFEDS sources \citep{Liu+2022}. Considering the continuation of this power-law, below $\sim \qty{1}{keV}$ a prominent soft excess component is present. This will be explored in greater detail in later sections.
Additionally, an iron line feature is observable around $\sim$ \qtyrange[range-phrase=--,range-units=single]{6}{7}{keV}. Such component, generally unexpected for single source spectrum due to \ero's poor effective area above $\sim\ \qty{2}{keV}$, appears in both the unbinned and rebinned ensemble stacked spectrum. This demonstrates the power of spectral stacking. This iron line feature cannot be attributed to the background, because the background iron lines \citep[e.g.,][]{Predehl+2021} arising at $z=0$ are blurred and dissolved as we rest-frame shift the source and background spectra.

\subsection{Differences in physical subgroups}\label{sec:pheno_data}
To investigate the variation of spectral profile with different physical properties, we divide the sample into three equally populated subgroups according to different physical quantities ($\alpha_\mathrm{ox}$, $L_\mathrm{UV}$\footnote{Note that for the $L_\mathrm{UV}$ division, we actually separate sources by the UV luminosity, $(\nu L_\nu)_\mathrm{UV}=\qty{2500}{\AA}\times L_\mathrm{UV}$.}, $\lambda_\mathrm{Edd}$, $M_\mathrm{BH}$), and stack sources in the same subgroup. The $\alpha_\mathrm{ox}$ and $L_\mathrm{UV}$ binnings are based on our ``spec-z'' sample, containing $\sim \num{5976}$ sources in each subgroup, while the $\lambda_\mathrm{Edd}$ and $M_\mathrm{BH}$ binnings are based on the ``BH-mass'' sample, containing $\sim \num{1390}$ sources in each subgroup. The resulting spectra normalized at \qty{4}{keV}, as shown in \cref{fig:pheno_data}, facilitates a pure data-driven analysis on the evolution of soft excess. We note that the $\alpha_\mathrm{ox}$- or $L_\mathrm{UV}$-binned spectra remain similar when stacking the ``BH-mass'' sample.  

We first uncover significant trends in \cref{fig:pheno_data} that the soft X-ray spectrum ($\lesssim 2\ \mathrm{keV}$) becomes stronger compared to hard X-ray for increasing $\alpha_\mathrm{ox}$ (first panel) and $\lambda_\mathrm{Edd}$ (third panel), aligning well with previous studies (\citealt{Walter&Fink1993,Grupe+2010}, and Fig. 14 from a similar stacking work, \citealt{Jin+2012a}), and extending these results to a larger population. In these two panels, the difference between stacked spectra is much larger than the uncertainties, which are often too small to be seen. For the $L_\mathrm{UV}$ binning (second panel), the trend is weaker, and a small difference is only apparent between the brightest subgroup and dimmest subgroup. In the right-most panel, the trend for $M_\mathrm{BH}$ is reversed. Less massive sources have steeper soft X-ray spectra than more massive sources, in agreement with previous studies \citep[e.g.][]{Porquet+2004,Piconcelli+2005}.

We present in \cref{fig:spec_diff} the difference spectrum between the highest $\alpha_\mathrm{ox}$ and lowest $\alpha_\mathrm{ox}$ subgroup from \cref{fig:pheno_data}.
The y-axis value indicates the strength of residual flux ($E^2N$) relative to hard X-ray power-law at \qty{4}{keV}. The significant residual in the soft X-ray (\qtyrange[range-phrase=--,range-units=single]{10}{100}{\%} that of hard X-ray power-law at \qty{4}{keV}) alongside the much weaker residual in the hard X-ray power-law, explicitly demonstrates that the soft excess cannot arise from reflection alone, unless the efficiency of hard X-ray power-law producing soft excess (e.g., the geometry) changes drastically with $\alpha_\mathrm{ox}$. The difference spectrum in \cref{fig:spec_diff} also reveals the intrinsic soft excess as a power-law between \num{0.3} and \qty{1}{keV}, where it starts to be ``cut off'', decreasing below the power-law by an e-fold at approximately \qty{2}{keV}.

Overall, the hard X-ray spectrum ($\gtrsim \qty{2}{keV}$) of \cref{fig:pheno_data} is more stable than the soft X-ray band. This is the case across different subgroups for all binnings, except that we find the lowest $M_\mathrm{BH}$ subgroup does seem to have the steepest hard X-ray power-law. However, the spectral quality in this regime is also relatively poor, inhibiting direct interpretation from data and requires further spectral fitting, which will be presented in the next section.

\subsection{Phenomenological spectral fits}
\label{sec:pheno_fit_results}

\begin{figure*}
    \centering
    \includegraphics[width=1.0\linewidth]{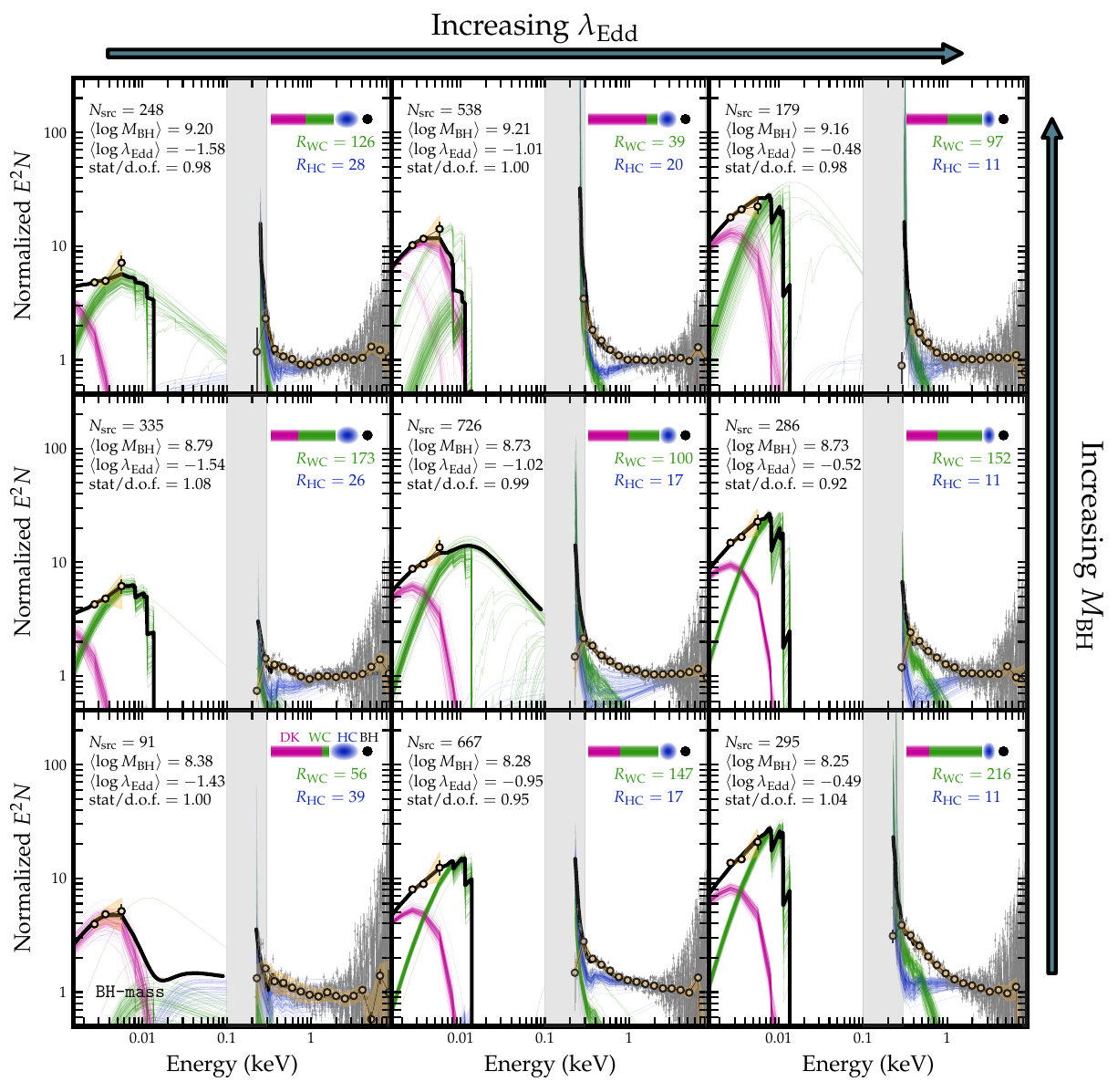}
    \caption{The optical-UV-Xray SED on our $3\times3$ $M_\mathrm{BH}$-$\lambda_\mathrm{Edd}$ grid, overlaid by best-fit bootstrap models, with red for disk (DK), green for warm corona (WC), blue for hot corona (HC). The legend gives the source number, median $M_\mathrm{BH}$, $\lambda_\mathrm{BH}$, best fitting statistic in each bin. A toy model illustrating the disk-corona size (warm corona radius $R_\mathrm{WC}$, hot corona radius $R_\mathrm{HC}$, both in units of $R_\mathrm{g}$) is shown on the top right of each panel. Note the potential star-formation contamination in two upper-left bins with low $\lambda_\mathrm{Edd}$ -- high $M_\mathrm{BH}$ when interpreting the results.}
    \label{fig:agnsed_model}
\end{figure*}

\begin{table*}
\caption{Spectral fitting results for the phenomenological model of different $\alpha_\mathrm{ox}/L_\mathrm{UV}/\lambda_\mathrm{Edd}/M_\mathrm{BH}$ binnings. The numbers in parenthesis for the subgroups indicate the median value of $\alpha_\mathrm{ox}/L_\mathrm{UV}/\lambda_\mathrm{Edd}/M_\mathrm{BH}$. We also list the number of sources ($N_\mathrm{src}$) and rest-frame \qtyrange[range-phrase=--,range-units=single]{0.2}{10}{keV} counts $C_\mathrm{0.2-10}$ in each subgroup.} \label{tab:specfit_tab1}
\resizebox{1.0\textwidth}{!}{
\centering
\begin{threeparttable}
\renewcommand{\arraystretch}{1.6}
\setlength{\tabcolsep}{3pt}
\begin{tabular}{c|cccccccccccc}
\toprule\toprule
Category & 
\multicolumn{3}{c}{$\alpha_\mathrm{oX}$} & 
\multicolumn{3}{c}{$\log L_\mathrm{UV}$} & 
\multicolumn{3}{c}{$\log\lambda_\mathrm{Edd}$} & 
\multicolumn{3}{c}{$\log M_\mathrm{BH}$} \\
\cmidrule(lr){1-1}
\cmidrule(lr){2-4}
\cmidrule(lr){5-7}
\cmidrule(lr){8-10}
\cmidrule(lr){11-13}
\multirow{2}{*}{Subgroup} & 
Low & Med & High & 
Low & Med & High & 
Low & Med & High & 
Low & Med & High \\ 
 & 
(1.25) & (1.41) & (1.56) & 
(44.58) & (45.31) & (45.84) & 
(-1.39) & (-0.93) & (-0.52) & 
(8.14) & (8.70) & (9.23) \\ 
\midrule
$q$ & 
$0.13_{-0.01}^{+0.01}$ & 
$0.27_{-0.02}^{+0.02}$ & 
$0.58_{-0.05}^{+0.06}$ & 
$0.27_{-0.02}^{+0.03}$ & 
$0.42_{-0.12}^{+0.06}$ & 
$0.40_{-0.05}^{+0.07}$\tnote{\dag} & 
$0.13_{-0.02}^{+0.02}$ & 
$0.33_{-0.07}^{+0.08}$ & 
$0.52_{-0.05}^{+0.07}$ & 
$0.32_{-0.08}^{+0.10}$ & 
$0.36_{-0.12}^{+0.10}$ & 
$0.25_{-0.06}^{+0.10}$\\ 
$E_\mathrm{cut}$ (keV) & 
$0.36_{-0.02}^{+0.04}$ & 
$0.62_{-0.05}^{+0.06}$ & 
$0.69_{-0.11}^{+0.06}$ & 
$0.63_{-0.05}^{+0.05}$ & 
$1.79_{-0.96}^{+0.21}$ & 
$2.00_{-0.00}^{+0.00}$\tnote{\dag} & 
$0.29_{-0.02}^{+0.05}$ & 
$0.89_{-0.18}^{+0.32}$ & 
$0.56_{-0.08}^{+0.07}$ & 
$0.78_{-0.17}^{+0.11}$ & 
$2.00_{-1.02}^{+0.00}$ & 
$0.30_{-0.03}^{+1.64}$\\ 
$\Gamma_\mathrm{SE}$ & 
$2.39_{-0.01}^{+0.02}$ & 
$2.42_{-0.02}^{+0.02}$ & 
$2.49_{-0.13}^{+0.26}$ & 
$2.45_{-0.02}^{+0.02}$ & 
$3.07_{-0.34}^{+0.17}$ & 
$3.20_{-0.17}^{+0.20}$\tnote{\dag} & 
$2.42_{-0.02}^{+0.02}$ & 
$2.53_{-0.15}^{+0.17}$ & 
$2.43_{-0.04}^{+0.09}$ & 
$2.56_{-0.06}^{+0.05}$ & 
$3.00_{-0.34}^{+0.24}$ & 
$2.47_{-0.02}^{+1.30}$\\ 
$\Gamma_\mathrm{PC}$ & 
$1.89_{-0.01}^{+0.02}$ & 
$1.91_{-0.02}^{+0.02}$ & 
$1.90_{-0.04}^{+0.04}$ & 
$1.95_{-0.02}^{+0.02}$ & 
$1.88_{-0.03}^{+0.03}$ & 
$1.87_{-0.02}^{+0.02}$\tnote{\dag} & 
$1.92_{-0.02}^{+0.02}$ & 
$1.87_{-0.05}^{+0.04}$ & 
$1.91_{-0.04}^{+0.04}$ & 
$2.06_{-0.06}^{+0.06}$ & 
$1.89_{-0.06}^{+0.05}$ & 
$1.95_{-0.04}^{+0.02}$\\ 
$n_\mathrm{H}\ (10^{20}\mathrm{cm}^{-2})$ & 
$1.05_{-1.05}^{+0.30}$ & 
$0.29_{-0.29}^{+0.28}$ & 
$1.72_{-1.72}^{+0.54}$ & 
$0.31_{-0.31}^{+0.15}$ & 
$1.24_{-1.24}^{+0.62}$ & 
$0.66_{-0.66}^{+1.98}$\tnote{\dag} & 
$2.19_{-0.55}^{+0.52}$ & 
$0.07_{-0.07}^{+0.68}$ & 
$1.02_{-1.02}^{+0.52}$ & 
$0.48_{-0.48}^{+0.48}$ & 
$1.69_{-0.74}^{+0.70}$ & 
$5.84_{-1.77}^{+1.61}$\\ 
PG stat/d.o.f. & 
$0.93$ & 
$1.10$ & 
$0.93$ & 
$1.08$ & 
$1.18$ & 
$1.07$\tnote{\dag} & 
$0.96$ & 
$1.00$ & 
$1.00$ & 
$0.96$ & 
$0.95$ & 
$0.96$\\ 
$N_\mathrm{src}$ & 
$5976$ & 
$5976$ & 
$5977$ & 
$5976$ & 
$5976$ & 
$5977$ & 
$1390$ & 
$1390$ & 
$1390$ & 
$1390$ & 
$1390$ & 
$1390$\\ 
$C_\mathrm{0.2-10}$ & 
\num{7e+05} & 
\num{7e+05} & 
\num{7e+05} & 
\num{8e+05} & 
\num{7e+05} & 
\num{6e+05} & 
\num{1e+05} & 
\num{1e+05} & 
\num{1e+05} & 
\num{1e+05} & 
\num{1e+05} & 
\num{1e+05}\\ 
\bottomrule
\end{tabular}
\begin{tablenotes}
\item[\dag]Ignoring 0.3 -- 0.4 keV due to very low source completeness ($<5\%$) in this energy range.
\end{tablenotes}
\end{threeparttable}
}
\end{table*}

Motivated by the apparent differences in spectra such as \cref{fig:spec_diff}, we now quantify the strength of the soft excess. We present in \cref{fig:pheno_model_alphaoXLUV} (for the $\alpha_\mathrm{ox}$ and $L_\mathrm{UV}$ binnings) and \cref{fig:pheno_model_EddMBH} (for the $\lambda_\mathrm{Edd}$ and $M_\mathrm{BH}$ binnings) the best-fit total model from a single fit (thick black line), as well as different best-fit components after 100 bootstrap experiments (green lines for soft excess, and blue lines for hard X-ray primary continuum). The distribution of the bootstrap realizations provide a clear view of parameter uncertainties as well as degeneracies. The source number, as well as fitting statistics are stated in each panel. Each panel also states the measured soft excess strength $q$. We use this to find the physical variable that causes the largest change in $q$. 

Overall, our phenomenological model yields an acceptable fit, mostly with the PG statistic per degree of freedom (d.o.f.) smaller than 1.20\footnote{As a rule of thumb, a $\mathrm{PG\ stat/d.o.f.}\sim1.0$ indicates a ``good'' fit, similar to the criterion often used for the traditional $\chi^2$ statistic. Our simulations show that for the sum of a series of Poissonian and Gaussian variables, the median of the resulting PG statistic distribution closely aligns with the value of $\mathrm{d.o.f.}$.}. For the ``High $L_\mathrm{UV}$'' subgroup, the fit is rather poor ($\mathrm{PG\ stat/d.o.f.}=1.32$), along with a high $n_\mathrm{H}$ value (pegged at the upper bound \qty{10e20}{cm^{-2}}). We consider this as a data problem arising from the relatively small number of sources effectively contributing to the \qtyrange[range-phrase=--,range-units=single]{0.3}{0.4}{keV} band, as the highest luminosity subgroup contains more high-redshift sources than the lower luminosity subgroups. Indeed, when ignoring \qtyrange[range-phrase=--,range-units=single]{0.3}{0.4}{keV}, the fitting statistic becomes much better (from 1.32 to 1.07) and $N_\mathrm{H}\sim\qty{0.66e20}{cm^{-2}}$, indicative of an unobscured spectrum, becomes more reasonable.

The strong variation of the soft excess strength $q$ and $\alpha_\mathrm{ox}$ or $\lambda_\mathrm{Edd}$, as shown in \cref{fig:pheno_data}, is now quantitatively confirmed. The soft excess is equivalent to approximately 10\% of the hard X-ray primary continuum flux in the \qtyrange[range-phrase=--,range-units=single]{0.5}{2.0}{keV} band at the lowest $\alpha_\mathrm{ox}$ or $\lambda_\mathrm{Edd}$ subgroup, increasing to approximately \numrange[range-phrase=--]{50}{60}\% at the highest $\alpha_\mathrm{ox}$ or $\lambda_\mathrm{Edd}$ subgroup. That is, the soft excess luminosity increases by a factor of five relative to the hard power-law. The trend is weaker for $L_\mathrm{UV}$ and $\log M_\mathrm{BH}$, with soft excess strength staying close to $\sim 0.30$ for each subgroup.

\Cref{tab:specfit_tab1} lists the spectral fit parameter values.
This includes their best-fit values as well as $68\%$ uncertainties from different physical subgroups of $E_\mathrm{cut}$, $\Gamma_\mathrm{SE}$, $\Gamma_\mathrm{PC}$ and $N_\mathrm{H}$. The median $\alpha_\mathrm{ox}$/$\log L_\mathrm{UV}$/$\log\lambda_\mathrm{Edd}$/$\log M_\mathrm{BH}$ for each subgroup are also included in parentheses on the third row. For $E_\mathrm{cut}$, we find a tentative positive correlation with $\alpha_\mathrm{ox}$, and a relatively strong correlation with $L_\mathrm{UV}$. Most subgroups require a relatively flat soft excess with $\Gamma_\mathrm{SE}\sim 2.6$\footnote{Note, that the photon index of $\sim 2.6$ is indeed harder than $3.2$ that we show in \cref{fig:spec_diff}, but this is not contradictory as most of the cut-off energy derived from spectral fitting is also lower than the ``turn-down'' point in \cref{fig:spec_diff} ($\sim 1\ \mathrm{keV}$).}, and no significant trend can be found neither except for the $L_\mathrm{UV}$ binning. We also do not find significant evolution for $\Gamma_\mathrm{PC}$, despite that the ``Low $M_\mathrm{BH}$'' subgroup exhibits a softer value ($2.06_{-0.06}^{+0.06}$) than average. Overall, the absorption is mild, typical for type 1 AGNs and consistent with our assumption. A further discussion on the above trends will be presented in \cref{sec:Discussion}.

\section{Multiwavelength modeling with \agnsed} \label{sec:agnsed_fit_results}
\begin{figure*}
    \centering
    \includegraphics[width=0.9\linewidth]{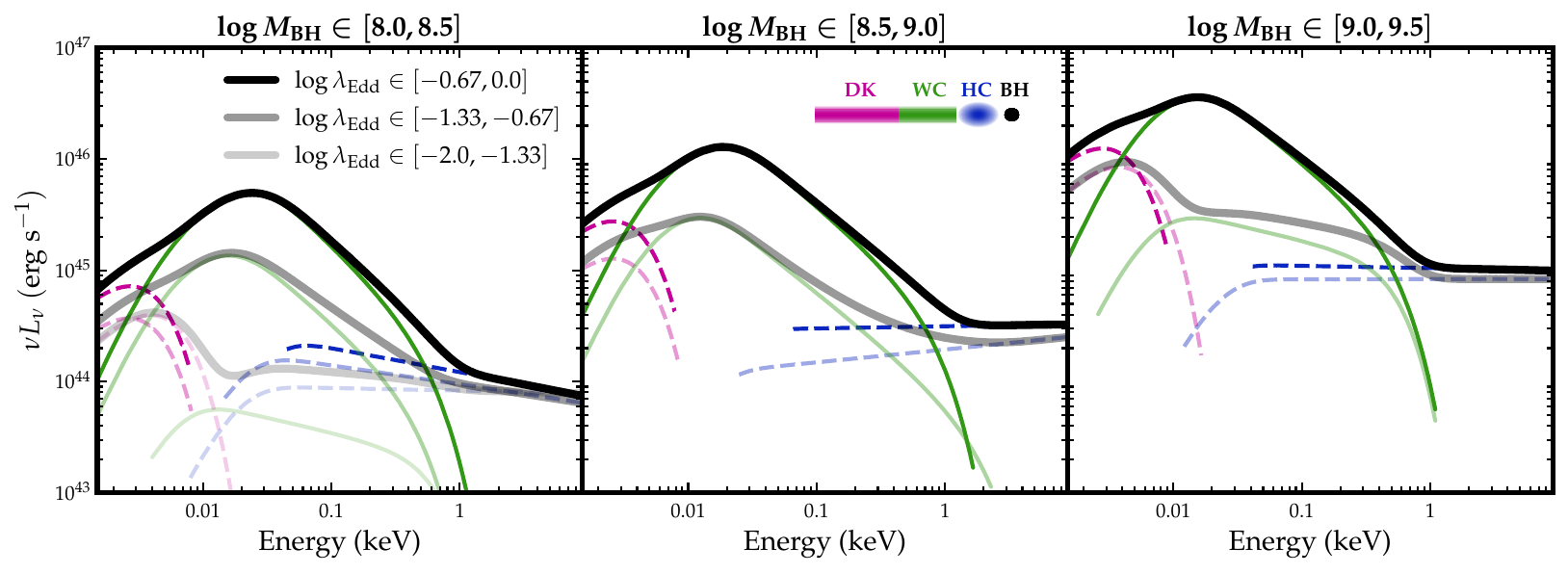}
    \caption{The best fit \agnsed~models for our $3\times 3$ $M_\mathrm{BH}$-$\lambda_\mathrm{Edd}$ grids (excluding the two with potential star formation contamination). Different components are shown, with the same color convention as in \cref{fig:agnsed_model}. As no data is involved, we scale the y-axis to physical values (\unit{erg.s^{-1}}), by setting \agnsed~distance parameter as $d_L=\qty{100}{Mpc}$, and normalization to $4\pi d_L^2$.}
    \label{fig:intrinsic_model}
\end{figure*}
\begin{figure*}
    \centering
    \includegraphics[width=1.0\linewidth]{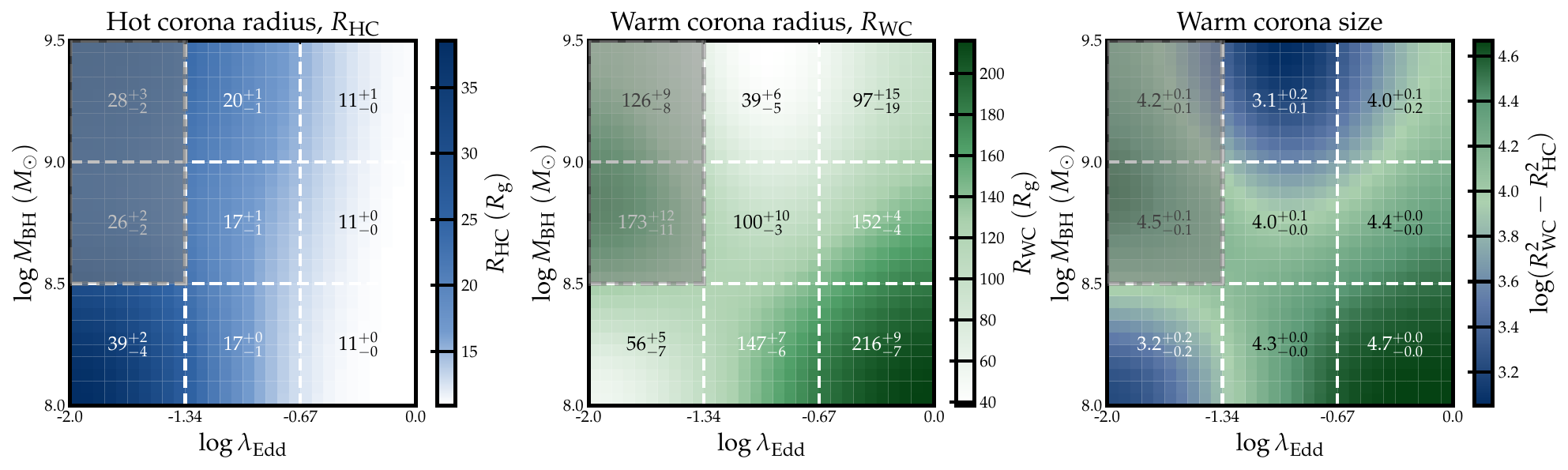}
    \caption{The evolutionary trend of hot corona radius $R_\mathrm{HC}$ (left), warm corona radius $R_\mathrm{WC}$ (middle), and warm corona ``size'' $\log (R_\mathrm{WC}^2-R_\mathrm{HC}^2)$ (right) on the $3\times3$ $M_\mathrm{BH}$-$\lambda_\mathrm{Edd}$ grid. The color is linearly interpolated for visualizing trends. The two low $\lambda_\mathrm{Edd}$ -- high $M_\mathrm{BH}$ bins are grayed out for star-forming contamination (see text for details), except for which the warm corona radius shows a clear increasing trend with increasing $\lambda_\mathrm{Edd}$ and decreasing $M_\mathrm{BH}$. The hot corona radius decreases with increasing $\lambda_\mathrm{Edd}$ while remains constant for different $M_\mathrm{BH}$. }
    \label{fig:RhcRwc}
\end{figure*}
\begin{table*}
\caption{The spectral fitting results for physical model (\agnsed). Additional scatter added to UV when doing bootstrapping.} \label{tab:specfit_tab2_new}
\footnotesize
\centering
\begin{threeparttable}
\renewcommand{\arraystretch}{1.6}
\setlength{\tabcolsep}{3pt}
\begin{tabular}{c|ccccccccc}
\toprule\toprule
BH-mass range & 
\multicolumn{3}{c}{$\log M_\mathrm{BH}\in[8.0,8.5]$} & 
\multicolumn{3}{c}{$\log M_\mathrm{BH}\in[8.5,9.0]$} & 
\multicolumn{3}{c}{$\log M_\mathrm{BH}\in[9.0,9.5]$} \\
\cmidrule(lr){1-1}
\cmidrule(lr){2-4}
\cmidrule(lr){5-7}
\cmidrule(lr){8-10}
$\langle \log M_\mathrm{BH}\rangle$ & 
8.38 & 8.28 & 8.25 & 
8.79 & 8.73 & 8.73 & 
9.20 & 9.21 & 9.16\\ 
$\langle \log\lambda_\mathrm{Edd}\rangle$ & 
-1.43 & -0.95 & -0.49 & 
-1.54 & -1.02 & -0.52 & 
-1.58 & -1.01 & -0.48\\ 
\midrule
$R_\mathrm{WC}\ (r_\mathrm{g})$ & 
$56_{-7}^{+5}$ & 
$147_{-6}^{+7}$ & 
$216_{-7}^{+9}$ & 
$173_{-11}^{+12}$ & 
$100_{-3}^{+10}$ & 
$152_{-4}^{+4}$ & 
$126_{-8}^{+9}$ & 
$39_{-5}^{+6}$ & 
$97_{-19}^{+15}$\tnote{\dag}\\ 
$R_\mathrm{HC}\ (r_\mathrm{g})$ & 
$39_{-4}^{+2}$ & 
$17_{-1}^{+0}$ & 
$11_{-0}^{+0}$ & 
$26_{-2}^{+2}$ & 
$17_{-1}^{+1}$ & 
$11_{-0}^{+0}$ & 
$28_{-2}^{+3}$ & 
$20_{-1}^{+1}$ & 
$11_{-0}^{+1}$\tnote{\dag}\\ 
$kT_\mathrm{WC}\ (\mathrm{keV})$ & 
$0.14_{-0.02}^{+0.03}$ & 
$0.12_{-0.02}^{+0.03}$ & 
$0.15_{-0.02}^{+0.03}$ & 
$0.12_{-0.02}^{+0.02}$ & 
$0.91_{-0.60}^{+0.09}$ & 
$0.21_{-0.04}^{+0.15}$ & 
$0.10_{-0.00}^{+0.01}$ & 
$0.12_{-0.01}^{+0.02}$ & 
$0.13_{-0.02}^{+0.03}$\tnote{\dag}\\ 
$\Gamma_\mathrm{WC}$ & 
$2.25_{-0.00}^{+0.00}$ & 
$2.94_{-0.10}^{+0.06}$ & 
$3.03_{-0.13}^{+0.09}$ & 
$2.50_{-0.06}^{+0.09}$ & 
$2.85_{-0.09}^{+0.07}$ & 
$2.94_{-0.14}^{+0.08}$ & 
$2.50_{-0.10}^{+0.08}$ & 
$2.25_{-0.00}^{+0.13}$ & 
$2.81_{-0.10}^{+0.13}$\tnote{\dag}\\ 
$\Gamma_\mathrm{HC}$ & 
$2.02_{-0.03}^{+0.03}$ & 
$2.16_{-0.04}^{+0.03}$ & 
$2.21_{-0.07}^{+0.06}$ & 
$1.96_{-0.06}^{+0.03}$ & 
$1.82_{-0.12}^{+0.13}$ & 
$1.98_{-0.17}^{+0.09}$ & 
$1.87_{-0.03}^{+0.03}$ & 
$1.99_{-0.04}^{+0.03}$ & 
$2.02_{-0.03}^{+0.05}$\tnote{\dag}\\ 
$n_\mathrm{H}\ (10^{20}\mathrm{cm}^{-2})$ & 
$0.21_{-0.21}^{+0.88}$ & 
$1.24_{-0.50}^{+0.58}$ & 
$3.64_{-0.84}^{+0.95}$ & 
$2.19_{-1.12}^{+0.74}$ & 
$0.29_{-0.29}^{+0.53}$ & 
$5.80_{-2.23}^{+2.46}$ & 
$0.93_{-0.93}^{+1.38}$ & 
$4.17_{-1.06}^{+0.96}$ & 
$5.36_{-3.40}^{+2.67}$\tnote{\dag}\\ 
stat/d.o.f. & 
$1.00$ & 
$0.95$ & 
$1.04$ & 
$1.08$ & 
$0.99$ & 
$0.92$ & 
$0.98$ & 
$1.00$ & 
$0.98$\tnote{\dag}\\ 
$N_\mathrm{src}$ & 
$91$ & 
$667$ & 
$295$ & 
$335$ & 
$726$ & 
$286$ & 
$248$ & 
$538$ & 
$179$\\ 
$C_\mathrm{0.2-10}$ & 
\num{1e+04} & 
\num{6e+04} & 
\num{3e+04} & 
\num{4e+04} & 
\num{7e+04} & 
\num{2e+04} & 
\num{2e+04} & 
\num{4e+04} & 
\num{2e+04}\\ 
\bottomrule
\end{tabular}
\begin{tablenotes}
\item[\dag]Ignoring 0.3 -- 0.4 keV due to very low source completeness ($<5\%$) in this energy range.
\end{tablenotes}
\end{threeparttable}
\end{table*}

Our spectral binning by either $M_\mathrm{BH}$ or $\lambda_\mathrm{Edd}$ (\cref{sec:pheno_fit_results}) may potentially introduce selection effects due to the flux-limited nature of our sample. Specifically, the lowest (highest) $\lambda_\mathrm{Edd}$ bins preferentially contain sources with the highest (lowest) $M_\mathrm{BH}$, as shown in our \cref{fig:MBHEdd_dist} and also discussed in Section 6.1 of \citealt{Jin+2012a}. Also physically, we would like to tease out how the accretion disk and warm corona change with $\lambda_\mathrm{Edd}$ and $M_\mathrm{BH}$ behind the observed trend of soft excess, using the physical \agnsed~model. We therefore construct the optical-UV-Xray SED for our BH-mass sample on the $3\times3$ $M_\mathrm{BH}$-$\lambda_\mathrm{Edd}$ grid (\cref{fig:MBHEdd_dist}), following the stacking method described in \cref{sec:UV_stack}.

In \cref{fig:agnsed_model}, we present the optical-UV-Xray SED on our $3\times3$ $M_\mathrm{BH}-\lambda_\mathrm{Edd}$ grid, overlaid by best-fit bootstrap folded models. The source intrinsic models (absorption corrected) are shown in \cref{fig:intrinsic_model}. Generally, a stronger soft excess ($q$) and a larger UV-to-Xray ratio ($\alpha_\mathrm{ox}$) can be seen along the x-axis (increasing $\lambda_\mathrm{Edd}$). Along the y-axis (increasing $M_\mathrm{BH}$), the soft excess tentatively becomes weaker, and no significant trend can be seen for the UV-to-Xray ratio. Interestingly, the \nuv-\u~slope also seems to be steeper for increasing $\lambda_\mathrm{Edd}$, and flatter for increasing $M_\mathrm{BH}$, except for the two low $\lambda_\mathrm{Edd}$ -- high $M_\mathrm{BH}$ (top two bins in the first column) with relatively strong star-forming contamination. We note that these UV trends appear, even when we assume a constant UV SED slope (rather than a $\lambda_\mathrm{Edd}$/$M_\mathrm{BH}$-dependent slope) for K-correction, and therefore should be genuine.

The physical meaning behind these trends in the model can also be considered. In the top right corner of each panel in \cref{fig:agnsed_model}, an illustration of the disk-corona model geometry is shown, where the transitional radii of $R_\mathrm{WC}$ and $R_\mathrm{HC}$ are scaled to real values. More quantitatively, we present in \cref{fig:RhcRwc} the evolutionary trend for $R_\mathrm{HC}$, $R_\mathrm{WC}$ and $\log (R_\mathrm{WC}^2-R_\mathrm{HC}^2)$ (the ``size'' of the warm corona relative to the hot corona), as a function of $\lambda_\mathrm{Edd}$ and $M_\mathrm{BH}$. We label the bootstrap median as well as $68\%$ range for each value. The color map is linearly interpolated for visualizing trends. The low $\lambda_\mathrm{Edd}$ -- high $M_\mathrm{BH}$ bins are grayed out for star-forming contamination (see \cref{sec:grid}). For the remaining bins, as $\lambda_\mathrm{Edd}$ increases, $R_\mathrm{HC}$ becomes smaller, and $R_\mathrm{WC}$ increases, indicating that the warm corona increases in relative size. On the other hand, as $M_\mathrm{BH}$ increases, $R_\mathrm{HC}$ remains relatively constant, while $R_\mathrm{WC}$ decreases, indicating a smaller warm corona.

The coronal radius evolution and our pure data-driven analysis of spectral shape are consistent. Since most of the X-ray photons are emitted from the hot corona, in the \agnsed{} model the hot corona radius reflects the relative importance of X-ray to total luminosity ($\sim\ \alpha_\mathrm{ox}$). The negative correlation between $R_\mathrm{HC}$ and $\lambda_\mathrm{Edd}$ simply manifests that $\alpha_\mathrm{ox}$ increases with $\lambda_\mathrm{Edd}$. The warm corona radius, on the other hand, is more complex and relies on both UV and soft X-ray. First, $R_\mathrm{WC}$ is related to the outer edge warm corona seed photon temperature, which peaks at UV and is determined from \nuv-\u~slope. Second, since a larger warm corona produces more soft X-ray emission relative to the hard X-ray primary continuum (hot corona), $R_\mathrm{WC}$ should also be related to the soft excess strength $q$, defined as the ratio of soft excess to hard X-ray primary continuum (in the most extreme case, if there is no soft ``excess'', then there should be no warm corona). In our case, the $R_\mathrm{WC}$ increase (decrease) with $\lambda_\mathrm{Edd}$ ($M_\mathrm{BH}$) is generally consistent with the observational trend that $q$ and \nuv-\u~slope increases (decreases or remains constant) with $\lambda_\mathrm{Edd}$ ($M_\mathrm{BH}$) in \cref{fig:agnsed_model}.

No significant trends are found aside from $R_\mathrm{HC}$ and $R_\mathrm{WC}$ (\cref{tab:specfit_tab2_new}). The warm corona photon index $\Gamma_\mathrm{WC}$ is found between $2.25\sim 3$, and tentatively becomes steeper for increasing $\lambda_\mathrm{Edd}$. For the hot corona photon index $\Gamma_\mathrm{HC}$, we find that its correlation with $\lambda_\mathrm{Edd}$ is rather weak for higher BH masses ($\log M_\mathrm{BH}\in[8.5,9.5]$). $\Gamma_\mathrm{HC}$ generally becomes larger for lower BH masses ($\log M_\mathrm{BH}\in [8.0,8.5]$), and a positive correlation between $\Gamma_\mathrm{HC}$ and $\lambda_\mathrm{Edd}$ is revealed. Finally, the majority of intrinsic absorption column density $n_\mathrm{H}$ consistently fall below \qty{10e20}{cm^{-2}}. We originally found the $n_\mathrm{H}$ at the highest $M_\mathrm{BH}$ -- highest $\lambda_\mathrm{Edd}$ (equivalently highest luminosity) bin pegged at \qty{10e20}{cm^{-2}}. For the same reason as \cref{sec:pheno_fit_results}, we ignore \qtyrange[range-phrase=--,range-units=single]{0.3}{0.4}{keV}, and find that $n_\mathrm{H}$ returns to a more realistic value of $\sim\ \qty{5e20}{cm^{-2}}$.

\section{Discussion} \label{sec:Discussion}
\begin{figure*}
    \centering
    \includegraphics[width=0.99\linewidth]{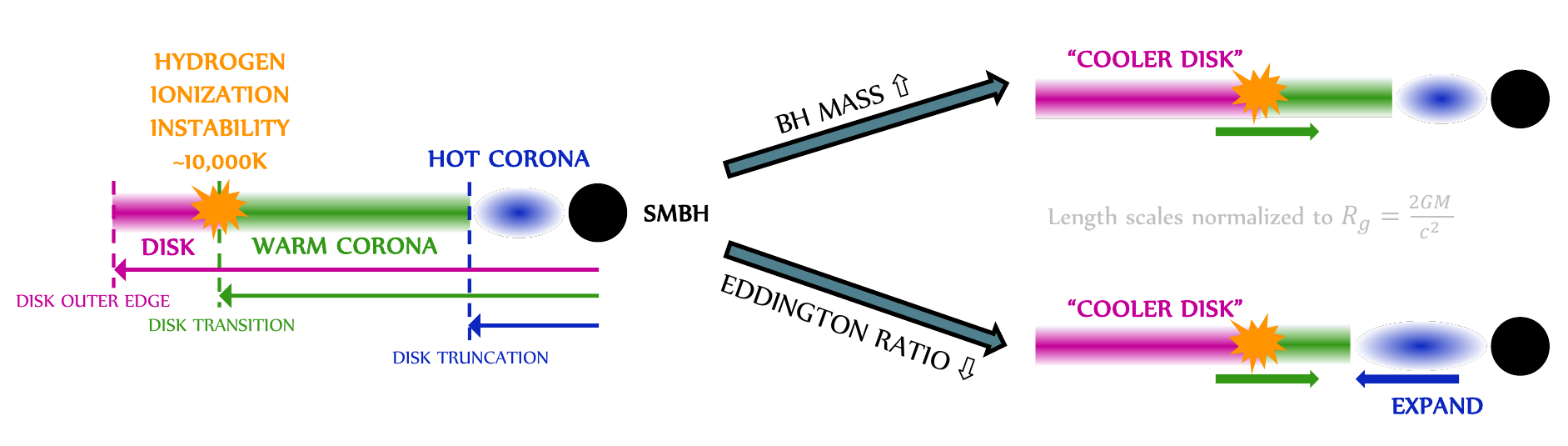}
    \caption{A sketch summarizing the physical picture of disk-to-warm corona transition. A hydrogen ionization instability at $\sim\qty{1e4}{K}$ within the disk likely triggers the formation of the warm corona. For either a more massive BH or a lower Eddington ratio, the disk temperature becomes cooler, causing the transition radius (in terms of $R_\mathrm{g}$) to move inward. Meanwhile, the hot corona expands (in units of $R_\mathrm{g}$) as Eddington ratio decreases, to match the observed increase in Xray-to-bolometric luminosity fraction.}
    \label{fig:sketch}
\end{figure*}

In this section, we compare our results with the literature and discuss the physical implications for the disk-corona connection. We first discuss some phenomenological properties of the soft excess and the relation with UV, for insights on the soft excess origin (\cref{sec:SE_UV}). Then, we focus on the disk radial structural change as implied by our data, starting from the standard outer disk to disk transition into warm corona (\cref{sec:disk_transition}), and further towards disk truncation into hot corona (\cref{sec:disk_truncation}), assuming the configuration of \agnsed. With our stacked optical-UV-Xray spectra, we also discuss in \cref{sec:BolCorr} the bolometric corrections as a function of many physical parameters. Finally, in \cref{sec:caveats}, we discuss some caveats of our study.

\subsection{The connection between UV and soft excess}\label{sec:SE_UV}
The physical origin of the soft excess has long been debated. In current understanding, either it originates from the inverse Compton scattering of disk UV photons \citep[e.g.,][]{Petrucci+2013,Petrucci+2018}, where it is more related to UV (disk) than hard X-ray (hot corona); or it could possibly be the blended reflection feature of hot corona illuminating on the inner ionized disk \citep[e.g.,][]{Ross&Fabian1993,Crummy+2005}, where it should show stronger dependence on the hard X-ray rather than the UV. Although both components can contribute to the real observed soft excess \citep[e.g.,][]{Chen+2025a}, the warm corona contribution has been proven to dominate by many recent studies, either by model comparison \citep[e.g.,][]{Waddell+2023} or directly measuring the fraction of warm corona to ionized reflection \citep[e.g.,][]{Ballantyne+2024}. 

We investigate the response of the soft excess by stacking X-ray spectra on groups of similar AGN parameters. Stacked spectra in groups of different $\alpha_\mathrm{ox}$ (left panel of \cref{fig:pheno_data}, top panel of \cref{fig:pheno_model_alphaoXLUV}) show that, as the disk becomes stronger compared to the hard X-ray, the soft excess responds accordingly while the hard X-ray spectral profile (both the hard X-ray power-law and the iron line at $\sim 6.4\ \mathrm{keV}$ which is indicative of the reflection feature) remains nearly unchanged. This trend supports the dominance of warm corona over ionized reflection in two ways. First, in the reflection scenario, the soft excess originates from hard X-ray reflecting off the inner accretion disk; a constant hard X-ray spectrum would therefore predict a relatively stable soft excess strength $q$, yet we now observe that a fivefold amplification of $q$ (\cref{fig:pheno_data}). Second, $\alpha_\mathrm{ox}$ quantifies the relative strength of the UV (where UV photons serve as seed photons for the warm corona) to X-ray, and we see that $q$ closely tracks the increase in $\alpha_\mathrm{ox}$. We note that the strong correlation between soft excess strength and $\alpha_\mathrm{ox}$ has also been reported in \rosat~AGN sample studies \citep[e.g.,][]{Walter&Fink1993}. Regarding $\alpha_\mathrm{ox}$ and hard X-ray photon index $\Gamma_\mathrm{PC}$, previous studies involving large AGN samples from \chandra~across a wide redshift range \numrange[range-phrase=--]{0.1}{4.7} \citep[e.g.,][]{Kelly+2007,Sobolewska+2009} also do not show a strong correlation. Several \xmm~studies of nearby bright Seyfert 1 galaxies do reveal a weak correlation between $\Gamma_\mathrm{PC}$ and $\alpha_\mathrm{ox}$ (\citealt{Jin+2012a}, and Chen et al., in prep). Whether the discrepancy arises from instrumental calibration (especially in the hard X-ray) is still unclear, and is postponed to a future study involving stacking \xmm~or \chandra~sources. Nevertheless, we note that the correlation reported in \citealt{Jin+2012a} is at most marginal ($p=0.01$), and from the three $\alpha_\mathrm{ox}$-binned stacked optical-UV-Xray SED (their Fig. 14) $\Gamma_\mathrm{PC}$ also changes only by $\sim\num{0.1}$. The correlation between $\Gamma_\mathrm{PC}$ and $\alpha_\mathrm{ox}$ in Chen et al. in prep is much weaker than that between soft X-ray slope and $\alpha_\mathrm{ox}$ (similar to this work). Therefore, our conclusions should remain largely consistent with previous studies.

In a flux-limited survey, $\alpha_\mathrm{ox}$ is correlated with $L_\mathrm{UV}$, so we should also anticipate a correlation between $q$ and $L_\mathrm{UV}$. As seen in \cref{fig:pheno_data} and lower panel of \cref{fig:pheno_model_alphaoXLUV}, such a trend does exist in our data, but is nevertheless weaker than the correlation between $q$ and $\alpha_\mathrm{ox}$. Therefore, it is possible that the soft excess strength is intrinsically related to the \textit{relative} power of UV in the total light, and the correlation between $q$ and absolute UV luminosity is just a secondary effect, due to the $\alpha_\mathrm{ox}\sim L_\mathrm{UV}$ relation. This might suggest that AGNs across different luminosities share the same X-ray spectral shape and can interestingly be linked to the finding that AGNs across different luminosities share the same optical-UV SED \citep[e.g.,][]{Cai&Wang2023a,Cai2024}. Despite this, we interestingly find in the stacked spectra that (\cref{tab:specfit_tab1}), as UV luminosity increases, the soft excess photon index $\Gamma_\mathrm{SE}$ increases (accompanied by an increase in cut-off energy $E_\mathrm{cut}$), while the hard X-ray photon index $\Gamma_\mathrm{PC}$ slightly decreases. This could possibly suggest stronger UV radiation enhances a competitive process between the soft excess component (warm corona) and the hot corona. That said, a parameter degeneracy test and a detailed treatment of the X-ray and optical selection bias are needed to validate this and will be carried out in a future study.

\subsection{The disk radial structural change: from disk to warm corona}\label{sec:disk_transition}

A Comptonized warm corona component can successfully explain the UV and soft excess, both in individual AGNs \citep[e.g.,][]{Kubota&Done2018} and stacked spectra (\citealt{Hagen+2024}, and our \cref{fig:agnsed_model}). However, how and why the disk transits into a warm Comptonized structure is not clearly understood. 

An interesting possibility is that such transition may relate to disk thermal instability near hydrogen ionization threshold ($\sim \qty{1e4}{K}$, e.g., \citealt{Hagen+2024}). Such concept (also known as ``Limit-Cycle Mechanism'' in literature), originally developed to account for periodic eruptions in dwarf novae \citep[e.g.,][]{Mineshige&Osaki1983,Cannizzo&Wheeler1984}, was later refined to explain AGN variability \citep[e.g.,][]{Cannizzo1992,Hameury+2009}. The basic idea is that the accretion disk hydrogen starts to be ionized at $\sim\qty{1e4}{K}$. A small increase in temperature leads to increase in ionization and opacity, trapping more heat and thereby further increases temperature; conversely, a small decrease in temperature reduces ionization and opacity, allowing the disk to cool further. This runaway behavior leaves the disk (at critical temperature) at two extreme states, and may oscillate in between. Such highly unstable pattern results in disk vertical structure very unlike that of the standard slim disk \citep{Cannizzo&Reiff1992}, and possibly be the key to inducing a warm corona. 

For a standard disk, the temperature $T_\mathrm{disk}$ at any given radius $R$ follows $T_\mathrm{disk}=\num{6.157e7}\times(\lambda_\mathrm{Edd}/M_\mathrm{BH})^{1/4}\eta_\mathrm{eff}^{-1/4}(R/R_\mathrm{g})^{-3/4}$ \citep[e.g.,][]{Shakura&Sunyaev1973,Novikov&Thorne1973}. Through \agnsed~modeling, our data reveals that the warm corona radius, $R_\mathrm{WC}$, increases with $\lambda_\mathrm{Edd}$ and decreases with $M_\mathrm{BH}$. Inserting the median $\lambda_\mathrm{Edd}$ and $M_\mathrm{BH}$ for each of the 7 grids free from star formation contamination, and assuming radiation efficiency factor $\eta_\mathrm{eff}=0.1$, we interesting find the disk temperature at the disk-to-warm-corona transitional radius roughly remains constant at $\sim \qty{1e4}{K}$, with $\sim \qty{0.3}{dex}$ of dispersion. We summarize this physical picture in \cref{fig:sketch}.

The constancy of transitional temperature is consistent with \citealt{Hagen+2024}, although our value ($\sim \qty{1e4}{K}$) is slightly higher than theirs ($\sim \qty{6e3}{K}$). While \qty{6e3}{K} may be physically related to the hydrogen ionization change \citep[e.g.,][]{Cannizzo&Reiff1992}, we note that a higher temperature ($\sim 10^4\ \mathrm{K}$) has also been found in many individual sources \citep[e.g.,][]{Kubota&Done2018}. Moreover, there could be systematic uncertainties affecting the specific value of such a temperature (e.g., intrinsic dust extinction, potential IGM absorption), and the standard disk model may be an incomplete description of AGN accretion \citep[e.g.,][]{Dexter&Agol2011,Netzer&Trakhtenbrot2014,Lawrence2018,Antonucci2018}. Therefore, more sophisticated treatment of the host galaxy and dust extinction is needed to validate the exact temperature in the future.

Despite the potential uncertainty on the transitional temperature, the systematic transition of the disk structure could be ubiquitous across all types of AGN. Recently, \citealt{Li+2024a} examined the disk structure of a changing-look AGN (\object{1ES 1927+654}, $M\approx10^6M_\odot$) at different epochs. They assume a two-zone structure, with a standard disk (producing UV/optical emission) in the outer region transiting\footnote{Note the slightly different terminology: the ``truncation'' refers to the separation between disk (warm corona) and hot corona in this work, but refers to the separation between the outer standard disk and inner slim disk in \citealt{Li+2024a}, which we will instead refer to as ``transition'').} into a slim disk (producing soft X-ray emission) in the inner radius, as well as a hot corona near the central BH (see their Fig. 1). This is not identical but comparable to the warm corona in our configuration. They found that the transitional radius increases at the high accretion state compared to low accretion state (see their Fig. 14), similar to the pattern observed in our \cref{fig:RhcRwc} and \citealt{Hagen+2024}. Comparing 1ES 1927+654 with other normal type 1 AGNs ($M\gtrsim10^7M_\odot$) at similar accretion rate, they also concluded that the transitional radius should decrease with increasing BH mass, similar to our findings in \cref{fig:RhcRwc}.

Regarding the warm corona temperature, the average we find, $0.22_{-0.06}^{+0.02}\ \mathrm{keV}$, is overall comparable to other literature results \citep[e.g.,][]{Petrucci+2018,Kubota&Done2018,Waddell+2023}. A slightly larger value found in \citealt{Petrucci+2018} and \citealt{Waddell+2023} could possibly be due to different energy bands or models used to constrain the warm corona. Similar to its proxy $E_\mathrm{cut}$, the warm corona temperature $kT_\mathrm{WC}$ does not exhibit significant correlation with $\lambda_\mathrm{Edd}$ or $M_\mathrm{BH}$. Our typical warm corona photon index $\Gamma_\mathrm{WC}$ is consistent with other literature \citep[e.g.,][]{Petrucci+2018}, and $\Gamma_\mathrm{WC}$ in general shows a non-monotonous correlation with $\lambda_\mathrm{Edd}$ or $M_\mathrm{BH}$. The lack of correlation between $kT_\mathrm{WC}$/$\Gamma_\mathrm{WC}$ with $\lambda_\mathrm{Edd}$/$M_\mathrm{BH}$ is in agreement with recent findings \citep[e.g.,][]{Waddell+2023,Palit+2024}. This could suggest that parameters additional to accretion rate and BH mass are controlling the warm corona, such as the magnetic field, \citealt{Gronkiewicz+2023}. Indeed, recent MHD simulation indicate the important role of magnetic field in warm corona formation as well as warm corona temperature \citep[e.g.,][]{Igarashi+2024}.

\subsection{The truncation of disk and the hot corona}\label{sec:disk_truncation}
The clear bend at $\sim$ \qtyrange[range-phrase=--,range-units=single]{1}{2}{keV} seen in both literature \citep[e.g.,][]{Boissay+2016} and our stacked spectra suggests the component at soft X-ray (warm corona) and hard X-ray cannot be the same structure with a continuous temperature gradient. \agnsed~assumes the disk (warm corona) truncates at a radius $R_\mathrm{HC}$ (without a physical motivation), below which the accretion flow turns into a hot Comptonized plasma, which contributes to the hard X-ray continuum.

Observationally, many works find that the hard X-ray continuum only contributes with a small fraction of the bolometric luminosity \citep[e.g.,][]{Marconi+2004,Kubota&Done2018,Duras+2020}, and such fraction (the ``X-ray loudness'') decreases with increasing accretion rate \citep[e.g.,][]{Vasudevan&Fabian2009,Ballantyne+2024}. This is also seen in \cref{fig:agnsed_model}, where with increasing $\lambda_\mathrm{Edd}$ the hard X-ray emission fraction becomes smaller in the normalized spectra. The X-ray to bolometric ratio is an important factor for constraining $R_\mathrm{HC}$ in \agnsed~(in that $R_\mathrm{HC}$ decreases for decreasing X-ray to bolometric ratio), which is why we see $R_\mathrm{HC}$ decreasing monotonously with $\lambda_\mathrm{Edd}$ in \cref{fig:RhcRwc}.

The physical driver behind the truncation of disk and $R_\mathrm{HC}$ (X-ray loudness) trend remains unclear. An interesting idea to explain the truncation of the disk involves the process of evaporation (see \citealt{Liu&Qiao2022} for review), a concept initially established for dwarf novae \citep[e.g.,][]{Meyer+2000} and later adapted to interpret the disk -- hot corona interaction in AGN \citep[e.g.,][]{Liu&Taam2009}. In this scenario, 
the hot corona is directly heated by viscous dissipation, and cools primarily via bremsstrahlung emission. Since the cooling efficiency of bremsstrahlung scales with the square of electron number density, a low-density hot corona cannot effectively radiate away its energy. Consequently, the outer layer of the disk, in contact with the hot corona, is heated, leading to the transfer of cooler gas into the corona to establish radiative equilibrium -- a process called ``disk evaporation''.
Disk truncation is thought to occur when the evaporation rate exceeds the accretion rate, with the truncation radius exhibiting a negative correlation with the Eddington ratio but no explicit dependence on BH mass, as suggested by numerical calculations \citep[][]{Taam+2012}. While the disk-corona geometry in the evaporation model differs from that assumed in \agnsed~(with the evaporation model placing the hot corona above the disk rather than radially segregated as in \agnsed), the observed anti-correlation between $R_\mathrm{HC}$ and $\lambda_\mathrm{Edd}$, as well as no correlation between $R_\mathrm{HC}$ and $M_\mathrm{BH}$ in \cref{fig:RhcRwc}, aligns qualitatively with the prediction of the evaporation model.

An alternative is that the hard X-ray photon index reflects the hot corona geometry to some extent. In our 1-dimension grid (\cref{tab:specfit_tab1}), we find that such value does not exhibit strong dependence on $\lambda_\mathrm{Edd}$ or $M_\mathrm{BH}$, except that the lowest-$M_\mathrm{BH}$ has the steepest photon index of \num{2.06}. Examining our $3\times3$ grid, we see that while $\Gamma_\mathrm{HC}$ correlates with $\lambda_\mathrm{Edd}$ at the lowest BH mass, the correlation disappears at higher BH mass, an observation consistent with \citealt{Laurenti+2024}. The intriguing divergence in $\Gamma_\mathrm{HC}$ trend around BH mass of $\sim \qty{1e8}{M_\odot}$ may suggest intrinsic differences in the innermost regions of AGNs responsible for the hard X-ray emission. It could either be that the high-mass AGNs have different hot corona geometry from that of low-mass AGNs, or that the jet contributes more strongly to the hard X-ray \citep[e.g.,][]{Kang&Wang2022} for them. Determining which of these scenarios is correct will require future work, particularly dedicated to the radio-loudness distribution in each bin.

\subsection{Bolometric corrections}\label{sec:BolCorr}
\begin{figure}
    \centering
    \includegraphics[width=1.0\linewidth]{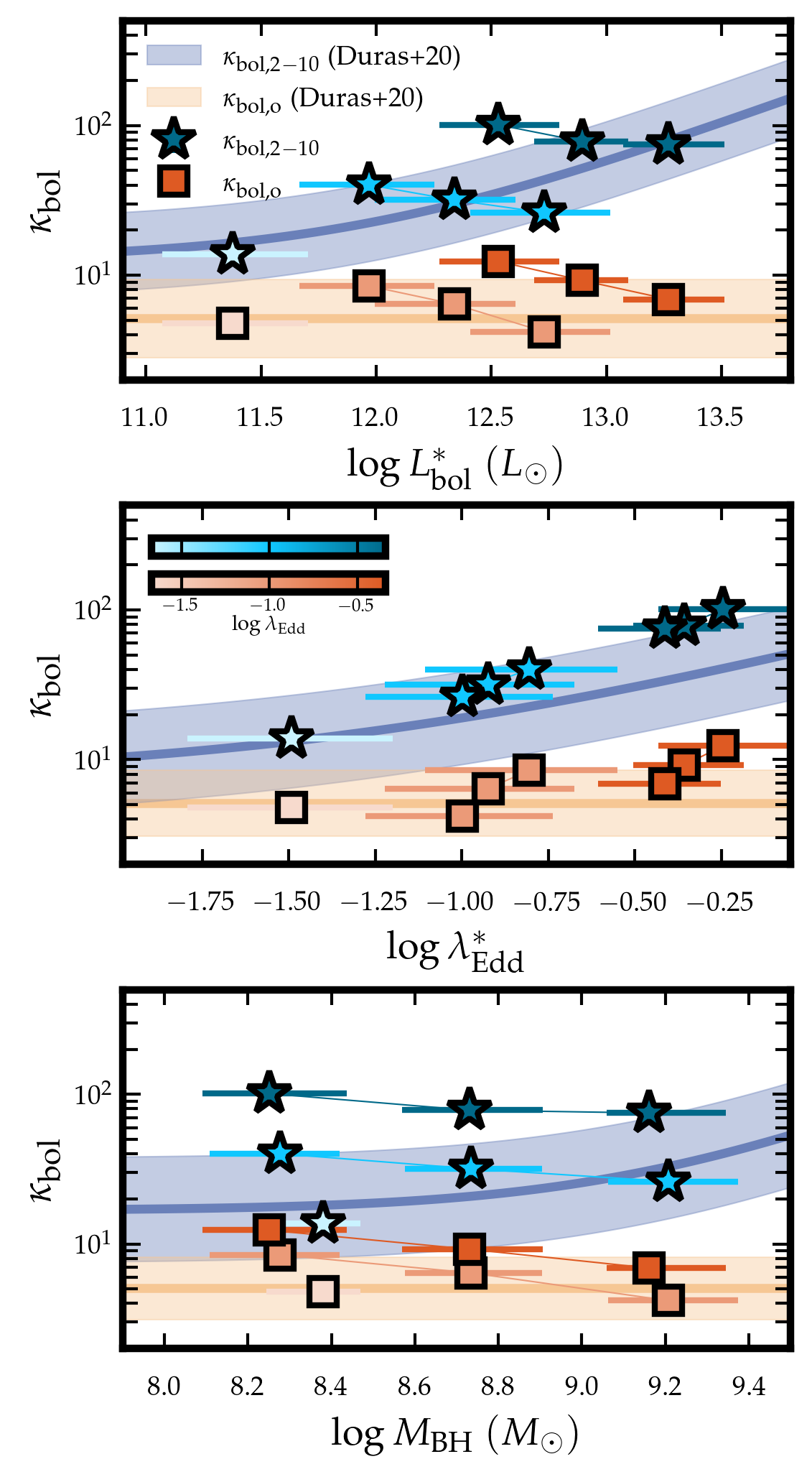}
    \caption{The B-band ($\kappa_\mathrm{bol,o}$, orange) and hard X-ray ($\kappa_\mathrm{bol,2-10}$, blue) bolometric correction factors for our stacked optical-UV-Xray SEDs, as a function of ``refined'' bolometric luminosity (top panel), ``refined'' Eddington ratio $\lambda_\mathrm{Edd}^*$ (middle panel), and BH-mass $M_\mathrm{BH}$ (bottom panel). Points with similar Eddington ratio ($\lambda_\mathrm{Edd}$, taken from \citealt{Wu&Shen2022}) are connected with lines and share the same color. The points overall straddle the literature relations from \citealt{Duras+2020}.}
    \label{fig:BolCorr}
\end{figure}
With our continuum spectral model from optical to hard X-ray we have a closer look at the bolometric emission. Constrained by both the NUV and soft X-ray, our physical model of the soft excess component bridges the difficult-to-observe FUV to extremely soft X-ray portion of the AGN spectrum, where substantial fractions of the luminosity is emitted. This improved accuracy allows us to use ``refined'' bolometric luminosities $L_\mathrm{bol}^*$ and Eddington ratios $\lambda_\mathrm{Edd}^*$ derived from our optical-UV-Xray model and the black hole masses listed in \citealt{Wu&Shen2022}. Thus, we do not use their bolometric luminosities and Eddington ratios, but recompute these. We treat each grid bin in \cref{fig:MBHEdd_dist} separately, skipping the two bins potentially contaminated by star-formation.
To obtain a per-source luminosity for each individual source, we refit the best-fit \agnsed~model (see \cref{sec:agnsed_fit_results}) from its corresponding $3\times 3$ grid bin, with only the normalization left free (for the same reason as in \cref{sec:fit_agnsed}). Then, from the absorption-corrected spectral model we then extract the hard X-ray luminosity (\qtyrange[range-phrase=--,range-units=single]{2}{10}{keV}, blue in \cref{fig:BolCorr}), the B-band luminosity (monochromatic luminosity at \qty{4400}{\AA} multiplied by \qty{4400}{\AA}, orange in \cref{fig:BolCorr}), as well as bolometric luminosity ($10^{-5}$ -- $10^3$ keV
). The \agnsed~model does not contain dust re-emission in the infrared, so counting from $10^{-5}$ to $10^3$ keV should safely incorporate all ``intrinsic'' emission from the accretion process \citep[see similar definition in e.g.,][]{Marconi+2004,Duras+2020}. Collecting these luminosities from each source in the same grid, we compute the median and $68\%$ confidence range.

In \cref{fig:BolCorr}, we present the X-ray bolometric correction factors in blue stars. The x-axis error bar of each grid point is in fact the 68\% ranges of individual sources in that grid. We do not plot error bars for the y-axis (the bolometric correction factors), since all sources in the same grid share the SED model with same shape (only normalization varies), and the scatter in bolometric correction factors among individual sources is very small. Points with similar $\lambda_\mathrm{Edd}$ are connected and share the same color, with lower $\lambda_\mathrm{Edd}$ having lighter colors. We also overlay the literature relations from \citealt{Duras+2020}. In the top panel of \cref{fig:BolCorr}, a positive correlation between $\kappa_\mathrm{bol,2-10}$ and $L_\mathrm{bol}^*$ can be observed, overall following the relation of \citealt{Duras+2020}. The correlation between $\kappa_\mathrm{bol,2-10}$ and $\lambda_\mathrm{Edd}^*$ is even tighter (middle panel), with only very small scatter in $\kappa_\mathrm{bol,2-10}$ at each fixed $\lambda_\mathrm{Edd}$. In contrast, the correlation between $\kappa_\mathrm{bol,2-10}$ and $M_\mathrm{BH}$ is very weak or even slightly reversed, at fixed $\lambda_\mathrm{Edd}$.

The optical bolometric correction factors, shown as red squares in \cref{fig:BolCorr}, are rather constant with bolometric luminosity. This is consistent with \citealt{Duras+2020}. A weak positive correlation with $\lambda_\mathrm{Edd}^*$ and anti-correlation with $M_\mathrm{BH}$ is observed.

\subsection{Caveats} \label{sec:caveats}
This worked aims to reveal the intrinsic soft X-ray spectra of AGNs above spectral transition ($\lambda_\mathrm{Edd}\gtrsim$ \numrange[range-phrase=--]{0.01}{0.02}), without alteration from absorption. To achieve this, we select unobscured AGN through optical-UV information, such as SDSS-confirmed type 1 broad-line AGNs in the ``BH-mass'' sample and retain in the ``spec-z'' sample only sources with similar $L_\mathrm{UV}$ and $L_\mathrm{X}$ as in the ``BH-mass'' sample. The resulting stacked spectra are smooth, with no significant absorption features. Spectral modeling consistently reveals at most weak neutral absorption ($n_\mathrm{H}\lesssim\qty{5e20}{cm^{-2}}$). 
Yet, warm absorption could still affecting the soft X-ray shape \citep[e.g.,][]{Waddell+2023,Nandra+2024}, given its presence even in ``unobscured'' type 1 AGNs \citep[e.g.,][]{Laha+2014}. The smoothness of our stacked spectra under \ero's good resolution of $\sim \qty{70}{eV}$ indicates that either warm absorption for our sample is overall weak, or the warm absorbers vary in velocity or ionization from source to source such that the effect is diluted when stacking. In the latter case, ideally obscured sources should be removed before stacking (see e.g., a Bayesian analysis by \citealt{Waddell+2023}), however the low photon counts of the \ero~all-sky sample makes this infeasible. Alternatively, systematic correlations between warm absorption and parameters used in this work ($L_\mathrm{UV}$, $\alpha_\mathrm{ox}$, $\lambda_\mathrm{Edd}$, $M_\mathrm{BH}$) could help quantify the fraction of absorbed flux in each stacked spectrum. This is left for future research. In this work, we model the stacked spectrum of the apparent soft excess as is with one choice of a physical model as a working hypothesis.

For the stacked optical-UV SED, we assume it is intrinsic to the accretion disk/warm corona and model it with \agnsed. However, recent AGN variability studies \citep[e.g.,][]{Korista&Goad2001,Korista&Goad2019,Netzer2022,Hagen+2024b} suggest that material around the broad-line region (BLR) can reprocess EUV/X-ray emission from the central engine and contribute additional optical-UV continuum emission (the so-called ``BLR diffuse continuum''). In addition, dust extinction and IGM absorption could also distort the optical-UV SED shape. A more careful treatment of these effects would likely require more \sdss~bands and stacked UV spectra, which we leave for future work.

For the soft X-ray excess, \agnsed~assumes that it is coming only from the warm corona. While the warm corona seems to play a dominant role in the soft X-ray regime, both argued by literature \citep[e.g.,][]{Hagen+2024} and evidenced in our \cref{fig:pheno_data}, a hybrid scenario where the ionized reflection is also contributing to a minor yet non-negligible portion of soft X-ray excess is favored by recent observations \citep[e.g.,][]{Ballantyne+2024,Chen+2025a} and theoretical work \citep[e.g.,][]{Ballantyne2020,Ballantyne&Xiang2020,Petrucci+2020}, especially when a high-density disk is considered \citep[e.g.,][]{Mallick+2025}. Future studies could test if incorporating the reflection components would affect the results. 

Another potential limitation of \agnsed~is its assumed geometry for the disk-corona system -- a series of concentric separated ``zones''. While this framework provides a useful simplification, it may not fully capture the complexity of the hot corona's structure. Recent \ixpe~polarimetric studies of black hole X-ray binaries and Seyfert 1 galaxies suggest that the hot corona likely has an extended ``slab-like'' geometry extended perpendicularly to the symmetry axis (and thus approximately along the equator; see e.g., \citealt{Gianolli+2023}). In this case, there could be certain overlap between the hot corona, the warm corona and the disk. Modeling this overlap would require more sophisticated models with additional free parameters to be designed, which may be too complex for the current data we have. For this reason, the simple geometry assumed by \agnsed~remains a reasonable and effective choice, allowing for a robust analysis. Additionally, the stacking technique we have developed lends itself to the application to the very large spectroscopic samples of \ero~selected AGN that will be delivered by SDSS-V and 4MOST in the coming years.


\section{Conclusion} \label{sec:conclusion}
The large number of low-counts X-ray spectra in \ero~era calls for spectral stacking for powerful scientific investigation. In this work we developed a new X-ray spectral stacking code, \xstack, which consistently stacks rest-frame PI spectra and response files with data-driven weighting factors. Compared with previous X-ray stacking codes \citep[e.g.,][]{Tanimura+2020,Tanimura+2022,Zhang+2024}, \xstack~offers advancements by optimally preserving X-ray spectral shape, incorporating Galactic absorption correction, and inheriting proper statistics for subsequent spectral fitting (\cref{sec:Xstack}).

Our X-ray spectral data comes from the \num{4.3} cumulative all-sky survey of \ero~(eRASS:5). This facilitates extremely high signal-to-noise ratio in our stacked spectra. Our ``spec-z'' sample, where all AGNs have reliable spectral redshift and \galex~\nuv~measurements, has a total exposure time of \qty{23}{Ms} and total photon counts within \qtyrange[range-phrase=--,range-units=single]{0.2}{10.0}{keV} of 2.1 million. Our ``BH-mass'' sample, where valid BH mass is further required, has total exposure time of \qty{3}{Ms} and total photon counts of \num{3.7e5} (\cref{sec:spec_extract}). With the \ero{} spectral resolution of $\sim \qty{70}{eV}$ in the soft band, we consistently retrieve smooth spectra absent of strong soft X-ray narrow lines.

We study how the \ero{} AGN X-ray spectra change as a function of $\alpha_\mathrm{ox}$, $L_\mathrm{UV}$, $\lambda_\mathrm{Edd}$, and $M_\mathrm{BH}$ (\cref{sec:pheno_data,sec:pheno_fit_results}). Then, further extending to UV and fitting with \agnsed~model on the $M_\mathrm{BH}-\lambda_\mathrm{Edd}$ grid, we investigate the physical properties of the warm/hot corona properties, with special focus on the coronal radius (\cref{sec:agnsed_fit_results}). Our main conclusions are:

\begin{enumerate}
    \item Phenomenologically, we find the soft X-ray ($\lesssim\qty{2}{keV}$) strength and spectral slope increases significantly with increasing $\alpha_\mathrm{ox}$, while the hard X-ray ($\gtrsim\qty{2}{keV}$) spectral shape remains nearly constant (\cref{fig:pheno_data} and \cref{tab:specfit_tab1}). This trend highlights soft excess’s dependence on the (relative strength of) UV emission while showing weaker correlation with hard X-rays, suggesting that soft excess is dominated by warm corona rather than ionized reflection (\cref{sec:SE_UV}). Similar trends are seen for the $\lambda_\mathrm{Edd}$ binning. The overall spectral shape evolution trend is weaker for $L_\mathrm{UV}$, and reversed for $M_\mathrm{BH}$. 

    \item We reveal the intrinsic ``soft excess'' spectral shape by subtracting the lowest $\alpha_\mathrm{ox}$ normalized stacked spectrum from the highest $\alpha_\mathrm{ox}$. The difference spectrum resembles a power-law below \qty{1}{keV}, with a cut-off above (\cref{fig:spec_diff}). This finding is consistent with the statement that the soft excess originates from a Compton-scattering dominated region (``warm corona'') with typical temperature $\lesssim\qty{1}{keV}$.

    \item Fitting our optical-UV-Xray stacked spectra with \agnsed~on the $3\times3$ $M_\mathrm{BH}$--$\lambda_\mathrm{Edd}$ grid (\cref{fig:MBHEdd_dist}), we find that the warm corona radius $R_\mathrm{WC}$ changes systematically with $\lambda_\mathrm{Edd}$ and $M_\mathrm{BH}$. $R_\mathrm{WC}$ generally increases with $\lambda_\mathrm{Edd}$ (except for the two low $\lambda_\mathrm{Edd}$ -- high $M_\mathrm{BH}$ bins potentially suffering from star formation contamination) while decreases with $M_\mathrm{BH}$ (\cref{fig:agnsed_model,fig:RhcRwc}). This suggests that the transition from standard disk to warm corona is triggered at a constant temperature \citep[proposed recently by][]{Hagen+2024}. The coronal radius trend remains robust even after controlling for sample selection biases (\cref{sec:robust_stack}). The warm corona temperature $kT_\mathrm{WC}$ and photon index $\Gamma_\mathrm{WC}$ are invariant (\cref{tab:specfit_tab2_new,sec:disk_transition}). 

    \item The hot corona radius $R_\mathrm{HC}$ decreases with $\lambda_\mathrm{Edd}$ while being insensitive to $M_\mathrm{BH}$. This can be explained by the disk evaporation model (\cref{sec:disk_truncation}). The hot corona photon index $\Gamma_\mathrm{HC}$ correlates primarily with $\lambda_\mathrm{Edd}$ at lowest $M_\mathrm{BH}$, while the correlation vanishes at higher $M_\mathrm{BH}$. Such intriguing trend may suggest different origins for the hard X-ray emission in AGN below/above $\sim \qty{1e8}{M_\odot}$.

    \item With our stacked optical-UV-Xray SED we also investigate the bolometric correction factors \cref{sec:BolCorr}. We find that the hard X-ray bolometric correction $\kappa_\mathrm{bol,2-10}$ increases with both bolometric luminosity and Eddington ratio, while shows weak or even slightly reversed relation with BH mass. For the optical B-band bolometric correction factor $\kappa_\mathrm{bol,o}$, our data reveals no evident relation with bolometric luminosity, a weak positive correlation with Eddington ratio, and a weak anti-correlation with BH mass.
    
\end{enumerate}

The exact transitional temperature still comes with large uncertainties, and further sophisticated treatment, fully accounting for host galaxy and dust/IGM extinction, which is now challenging for our \sdss~sample, will be carried out in the future.

\section*{Acknowledgements}
This work was supported by National Natural Science Foundation of China (Grant No. 124B1007) and USTC Fellowship (Grant No. U19582024).

We extend great appreciation to the referee for the constructive suggestions and invaluable feedback. We sincerely acknowledge Pietro Baldini from MPE for scientific discussions, Arne Rau from MPE for suggestion on individual source analysis, Konrad Dennerl, Tom Dwelly from MPE for help on \ero~technical issues, Joel Gil Leon from MPE for help on SciServer, Xiaoyuan Zhang, Xueying Zheng, Johan Comparat, Carolina Andonie and Soumya Shreeram from MPE, Victoria Toptun from The European Southern Observatory (ESO), Claudio Ricci from Universidad Diego Portales (UDP) for suggestions on X-ray spectral stacking, Luis Ho from Kavli Institute for Astronomy and Astrophysics at Peking University (KIAA-PKU) on discussion of UV data handling, Hao-Nan Yang, Raphael Shirley, William Roster, Gao-Xiang Jin from MPE, Steven Hämmerich from Friedrich-Alexander-Universität Erlangen-Nürnberg (FAU), Mirko Krumpe from Leibniz Institute for Astrophysics Potsdam (AIP), and Honghui Liu, Denys Malyshev from University of Tübingen (IAAT), Maria Chira from National Observatory of Athens (NOA), Yunliang Zheng from Shanghai Jiaotong University (SJTU), Wentao Lu from USTC for other helpful discussions.

\textit{Services:} This research has made use of the SIMBAD database \cite{Wenger+2000} and the VizieR catalog access tool \cite{Ochsenbein+2000}, operated at CDS, Strasbourg, France.
For bibliography, this research has made use of NASA's Astrophysics Data System, along with the adstex bibliography tool (\url{https://github.com/yymao/adstex}).

\textit{Software:} astropy \citep{AstropyCollaboration+2013,AstropyCollaboration+2018}, topcat \citep{Taylor2005}, stilts (\url{https://www.star.bris.ac.uk}), Sciserver \citep[implemented at MPE, following]{Taghizadeh-Popp+2020},  ultranest \citep{Buchner2021}, matplotlib \citep{Hunter2007}, scipy \citep{Jones+2001}, numpy \citep{Harris+2020}, numba \citep{Lam+2015}, joblib \citep{JoblibDevelopmentTeam2020}, GNU parallel \citep{Tange2011}, sfdmap (\url{https://github.com/kbarbary/sfdmap}).

\textit{X-ray spectroscopy survey data:}
This work is based on data from eROSITA, the soft X-ray instrument aboard SRG, a joint Russian-German science mission supported by the Russian Space Agency (Roskosmos), in the interests of the Russian Academy of Sciences represented by its Space Research Institute (IKI), and the Deutsches Zentrum für Luft- und Raumfahrt (DLR). The SRG spacecraft was built by Lavochkin Association (NPOL) and its subcontractors, and is operated by NPOL with support from the Max Planck Institute for Extraterrestrial Physics (MPE).
The development and construction of the eROSITA X-ray instrument was led by MPE, with contributions from the Dr. Karl Remeis Observatory Bamberg \& ECAP (FAU Erlangen-Nuernberg), the University of Hamburg Observatory, the Leibniz Institute for Astrophysics Potsdam (AIP), and the Institute for Astronomy and Astrophysics of the University of Tübingen, with the support of DLR and the Max Planck Society. The Argelander Institute for Astronomy of the University of Bonn and the Ludwig Maximilians Universität Munich also participated in the science preparation for eROSITA.
The eROSITA data shown here were processed using the eSASS/NRTA software system developed by the German eROSITA consortium.

\textit{Optical-UV spectroscopy survey data:} Funding for the Sloan Digital Sky Survey (SDSS) has been provided by the Alfred P. Sloan Foundation, the Participating Institutions, the National Aeronautics and Space Administration, the National Science Foundation, the US Department of Energy, the Japanese Monbukagakusho, and the Max Planck Society. The SDSS Web site is http://www.sdss.org/. 
The SDSS is managed by the Astrophysical Research Consortium (ARC) for the Participating Institutions. The Participating Institutions are The University of Chicago, Fermilab, the Institute for Advanced Study, the Japan Participation Group, The Johns Hopkins University, Los Alamos National Laboratory, the Max-Planck-Institute for Astronomy (MPIA), the Max-Planck-Institute for Astrophysics (MPA), New Mexico State University, University of Pittsburgh, Princeton University, the United States Naval Observatory, and the University of Washington.

\textit{Optical-UV photometric survey data:} The Legacy Surveys consist of three individual and complementary projects: the Dark Energy Camera Legacy Survey (DECaLS; Proposal ID \#2014B-0404; PIs: David Schlegel and Arjun Dey), the Beijing-Arizona Sky Survey (BASS; NOAO Prop. ID \#2015A-0801; PIs: Zhou Xu and Xiaohui Fan), and the Mayall z-band Legacy Survey (MzLS; Prop. ID \#2016A0453; PI: Arjun Dey). DECaLS, BASS and MzLS together include data obtained, respectively, at the Blanco telescope, Cerro Tololo Inter-American Observatory, NSF’s NOIRLab; the Bok telescope, Steward Observatory, University of Arizona; and the Mayall telescope, Kitt Peak National Observatory, NOIRLab. Pipeline processing and analyses of the data were supported by NOIRLab and the Lawrence Berkeley National Laboratory (LBNL). The Legacy Surveys project is honored to be permitted to conduct astronomical research on Iolkam Du’ag (Kitt Peak), a mountain with particular significance to the Tohono O’odham Nation.
NOIRLab is operated by the Association of Universities for Research in Astronomy (AURA) under a cooperative agreement with the National Science Foundation. LBNL is managed by the Regents of the University of California under contract to the U.S. Department of Energy.
The Legacy Survey team makes use of data products from the Near-Earth Object Wide-field Infrared Survey Explorer (NEOWISE), which is a project of the Jet Propulsion Laboratory/California Institute of Technology. NEOWISE is funded by the National Aeronautics and Space Administration. The Legacy Surveys imaging of the DESI footprint is supported by the Director, Office of Science, Office of High Energy Physics of the U.S. Department of Energy under Contract No. DE-AC02-05CH1123, by the National Energy Research Scientific Computing Center, a DOE Office of Science User Facility under the same contract, and by the U.S. National Science Foundation, Division of Astronomical Sciences under Contract No. AST-0950945 to NOAO.
GALEX (Galaxy Evolution Explorer) is a NASA Small Explorer, launched in April 2003. We gratefully acknowledge NASA’s support for construction, operation, and science analysis for the GALEX mission, developed in cooperation with the Centre National d’Etudes Spatiales (CNES) of France and the Korean Ministry of Science and Technology.

\section{Code \& data availability}
Our X-ray spectral stacking code, \xstack, will be made publicly available at \url{https://github.com/AstroChensj/Xstack}. Alongside the code, we also release our BH-mass sample, comprising \num{4170} eRASS-SDSS broad-line AGNs with measurements of BH mass, Eddington ratio, and X-ray luminosity. Since the full eRASS:5 dataset is scheduled for release in summer 2028, the X-ray luminosities provided here are based on eRASS1 data instead. The final stacked eRASS:5 spectra, under different physical binnings, will also be made available through Vizier.

\bibliographystyle{aa}
\bibliography{SE_eRO}

\appendix
\section{Detailed illustration of our X-ray spectral stacking code}\label{sec:Xstack}
In this section, we present a detailed illustration of our X-ray spectral stacking code, \xstack. We first review several basic concepts of X-ray spectroscopy, as well as the motivation of our code in \cref{sec:basic_concepts}. Then in \cref{sec:spec_shift,sec:spec_stack}, we introduce the specific process of spectral shifting and stacking. We finally summarize the code implementation and discuss some technical considerations in \cref{sec:xstack_notes}.

\subsection{Basic concepts}\label{sec:basic_concepts}
Suppose an X-ray instrument is observing an astronomical source that emits a specific photon spectrum (\unit{Photons.cm^{-2}.s^{-1}.keV^{-1}}). At any given time and energy, the number of X-ray photons (\unit{Photons.s^{-1}.keV^{-1}}) as seen by the CCD after the mirror reflection and filter absorption, is derived as the multiplication of input photon spectrum and effective area (\unit{cm^2}), with the latter varying with energy and being encoded in the ARF. Inside the CCD and during the readout, a complex electronic process takes place to convert the input photons (\unit{Photons.s^{-1}.keV^{-1}}) into calibrated events in each output energy channel (\unit{Counts.s^{-1}.keV^{-1}}), or PI spectrum, which is what we finally observe. The probability that a photon with some input energy falls in some energy channel is characterized by the RMF matrix, and the final number of events we see in each channel is described by a Poisson realization. The spectral resolution can be roughly estimated from the full width of half maximum (FWHM) of the RMF matrix diagonal bands.

Mathematically, the equation connecting the input model spectrum, response, background, and output PI spectrum, can be written as:
\begin{align}
    C_i&\sim \mathrm{Poisson}\left(\lambda_i+b_i\right)\\
    &=\mathrm{Poisson}\left(\sum_j R_{ij}\times A_{j}\times F_{j}\times \Delta t\times \Delta E_{j}+b_i\right)\label{eq:basic_separate} \\
    &=\mathrm{Poisson}\left(\sum_j \mathbf{P}_{ij}\times F_j\times\Delta t\times \Delta E_j+b_i\right)\label{eq:basic}
\end{align}
Here, $F_{j}$ is the spectral model flux in input model energy $E_{j}$ (\unit{Photons.cm^{-2}.s^{-1}.keV^{-1}}), $\Delta t$ is exposure time, $\Delta E_{j}$ is energy width, and $b_i$ is the predicted background contribution in channel $E_{i}$. The ARF $A_{j}$ encodes the instrumental effective area (\unit{cm^2}) to photons with model energy $E_{j}$, while RMF matrix $R_{ij}$ characterizes the probability of input model energy $E_{j}$ being reconstructed as output channel energy $E_{i}$. 
Integrating $\mathbf{P}_{ij}\times \Delta t\times \Delta E_{j}$ over the entire input energy space gives $\lambda_i$, the expected counts in output energy channel $E_{i}$. Finally, after the Poisson realization, $C_{i}$ is the spectral PI we observe. We refer readers to \citealt{Fioretti&Bulgarelli2020} and \citealt{Buchner&Boorman2023a} for more details on this topic. The standard formats for PI spectrum and RMF/ARF are documented by HEASARC\footnote{\url{https://heasarc.gsfc.nasa.gov/docs/heasarc/ofwg/docs/spectra/ogip_92_007.pdf} for PHA/PI and \url{https://heasarc.gsfc.nasa.gov/docs/heasarc/caldb/docs/memos/cal_gen_92_002/cal_gen_92_002.pdf} for RMF/ARF.}.

A common goal for X-ray spectroscopy is to infer the input spectral model, $F_{j}$, for each source. In the context of low-count X-ray spectroscopy, however, such task is challenging and comes with large uncertainty. A modern statistical approach is hierarchical Bayesian modeling \cite[see, e.g.,][for applications in X-ray spectroscopy]{Baronchelli+2018,Burgess2019}, which requires assuming a suitable model for each source, allowing each source to have different spectral parameters, and learning these in addition to the spectral parameter distribution. While powerful, hierarchical Bayesian modeling can be difficult to compute numerically, and can be unstable in the case (as applicable here), where each source contributes with very little information. Furthermore, they can be sensitive to outliers and artifacts. A more classical approach is the stacking (summation) of low-count spectra. A caveat of stacking is that the stacked spectrum is a summation of a potentially heterogeneous sample, and thereby may not necessarily represent the spectrum of any one source, even if observed for a long time. This criticism however also applies to time-averaged spectra of variable sources, which are commonly used in spectral fitting. A major benefit of spectral stacking is that the data can be visualized without requiring an a-priori spectral model, and thus allowing a data-driven interpretation of averaged spectral properties.

Despite being classical, stacking X-ray comes with its own challenges as compared to stacking in other bands (e.g., optical). First of all, X-ray typically lies in a low-counts regime and follows Poisson statistics (contrary to optical, which typically lies in a high-counts regime and can then be approximated with Gaussian statistics), meaning that the X-ray spectral counts and uncertainties cannot be scaled simultaneously. While in optical it is a common practice to first scale all spectra to similar flux level and scale the uncertainty accordingly, in X-ray this is generally not possible, as the uncertainty for the scaled X-ray spectrum cannot be trivially calculated (e.g., dividing a Gaussian variable $\mathcal{N}(\mu,\sigma^2)$ by \num{2} still results in a Gaussian $\mathcal{N}(\mu/2,\sigma^2/4)$, but dividing a Poisson variable $\mathrm{Poisson}(\lambda)$ by \num{2} does not result in $\mathrm{Poisson}(\lambda/2)$). Secondly, X-ray has poor spectral resolution ($R\sim\numrange{10}{100}$) and complex response, meaning that the input model energy and output channel energy is far from being on a 1:1 correlation (unlike the optical, where the spectral resolution is much better, $R\sim 10^3$). Therefore, the responses must also be taken into account when stacking, aside from the spectra.

A spectrum is characterized by its spectral shape and normalization. However, due to the Poisson nature of X-ray photons and the complexity of instrumental responses, preserving both properties in a stacked spectrum is challenging. In this work, we focus on maintaining the spectral shape. For this purpose, proper weights must be assigned to the responses during the stacking (see \cref{sec:spec_stack}). Crucially, we remark that any weighting factor must be assigned to the \textit{full response} $\mathbf{P}_{ij}$ rather than the RMF and ARF separately (the latter inappropriate approach having been commonly adopted in previous works). The reason is that RMF and ARF are combined multiplicatively in \cref{eq:basic_separate}. 

Additionally, for extragalactic sources, an accurate correction for Galactic absorption must be taken into account.

In an effort to tackle these issues, we develop \xstack, an open-source code for X-ray spectral stacking. Our methodology is to first sum all (rest-frame) PI spectra, without any scaling; and then sum the rest-frame, Galactic-$N_\mathrm{H}$ corrected full response, each with appropriate weighting factors to preserve the overall spectral shape. In the following sections, we provide step-by-step illustrations for the stacking procedure.

Throughout the following sections, we assume all sources come from the same instrument, i.e., they have same output channel energy edges (specified in \texttt{EBOUNDS} extension in RMF) and same input model energy edges (specified in \texttt{SPECRESP} extension in ARF). The rest-frame corrections also do not alter these edges.

\subsection{Spectral shifting} \label{sec:spec_shift}
\begin{figure}
    \centering
    \includegraphics[width=1.0\linewidth]{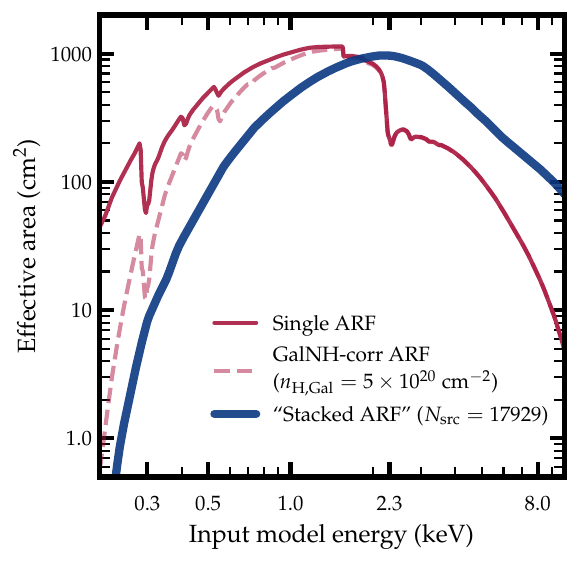}
    \caption{A figure comparing the single ARF (red solid) and the ``stacked ARF'' (blue solid) extracted from the stacked full response of the spec-z sample. We also demonstrate the way we correct for Galactic absorption for each source, that is, taking an energy-bin-wise product between the single ARF and Galactic absorption profile (red dashed), with the column density fixed at the value corresponding to that source.}
    \label{fig:ARF_plots}
\end{figure}
\begin{figure}
    \centering
    \includegraphics[width=0.9\linewidth]{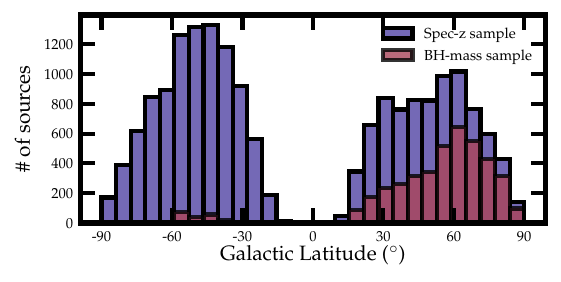}
    \caption{The Galactic latitude distribution of our spec-z and BH-mass sample. $\gtrsim90\%$ of sources in both samples have relatively high Galactic latitude ($|l|>\ang{30;;}$), where the Galactic medium is less clumpy and the estimation of column density from \nh~tool is more reliable.}
    \label{fig:bii}
\end{figure}
\begin{figure}
    \centering
    \includegraphics[width=0.9\linewidth]{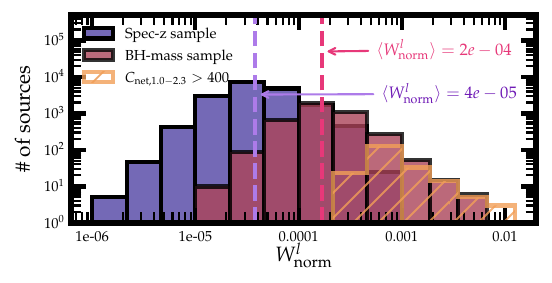}
    \caption{Distribution of normalized response weighting factor. As in \cref{fig:zcountdist}, the spec-z sample (purple filled), the BH-mass (red filled), and over-bright sources in spec-z sample (orange hatched) are shown, as well as median values with vertical dashed lines. The weighting factors, a combination of source counts and effective area in the \qtyrange[range-phrase=--,range-units=single]{1.0}{2.3}{keV} range, are distributed over several orders of magnitude.}
    \label{fig:RSP_weight}
\end{figure}
\begin{figure*}
    \centering
    \includegraphics[width=1.0\linewidth]{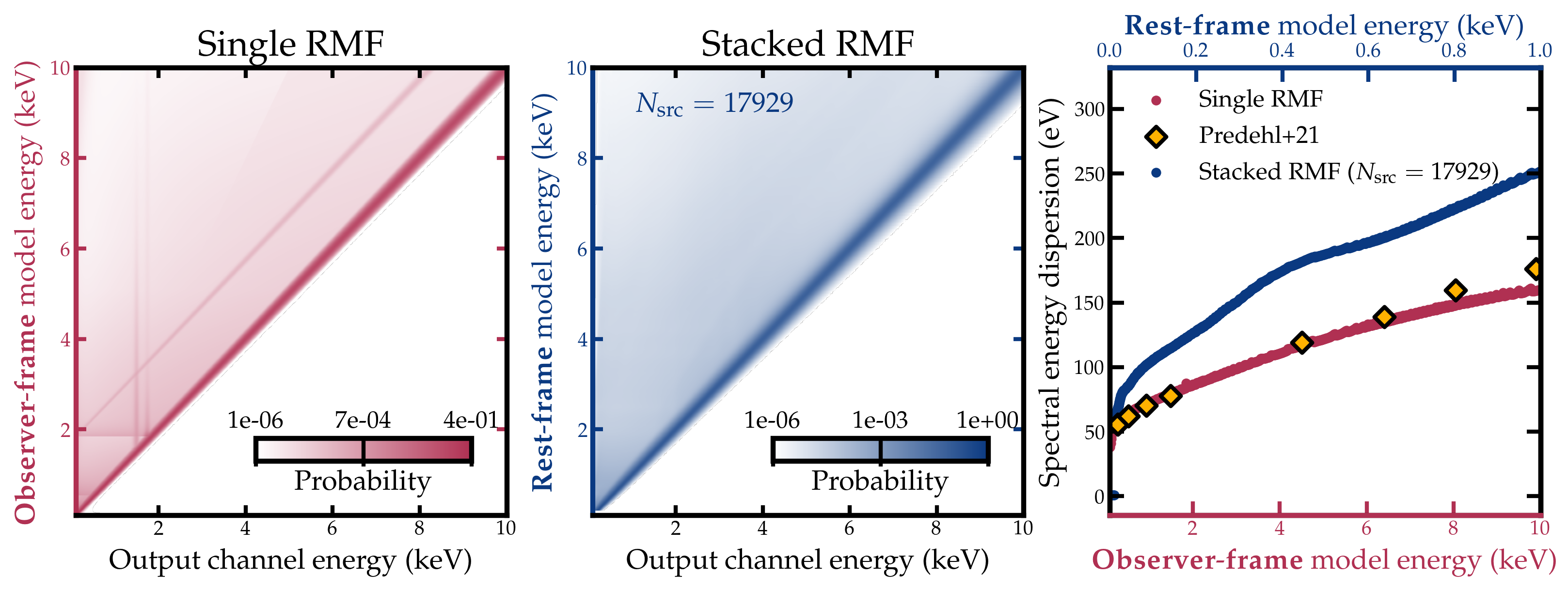}
    \caption{Effect of stacking RMFs. \textit{Left panel}: The observer-frame \ero{} RMF of a single source, where the color indicates the probability. Weak off-diagonals are visible. \textit{Middle panel}: rest-frame ``stacked RMF'' extracted from the stacked full response of the spec-z sample. Off-diagonals are less prominent, while the width of the diagonal becomes larger compared with a single RMF. \textit{Right panel}: spectral energy dispersion (FWHM of the RMF matrix diagonal) at different input model energies (observer-frame for the single RMF, while rest-frame for the stacked RMF), quantifying the spectral resolution. The energy dispersion increases with energy (resolution decreases with energy). The observer-frame energy dispersion of a single source (red curve) is lower than the rest-frame energy dispersion of the spec-z sample (blue). The standard measurement of \ero~energy resolution from Table 4 of \citealt{Predehl+2021} (7 TMs averaged) is also listed as orange diamonds.}
    \label{fig:RMF_plots}
\end{figure*}
Before stacking, we need to shift all sources to the common rest-frame, given the redshift of each source. The PI spectrum is essentially a histogram, with photon counts distributed across the output energy channels. For each energy channel $E\sim E+\Delta E$ in a source with redshift $z$, we redistribute its photon counts to the energy range $(1+z)E\sim (1+z)(E+\Delta E)$. Since the energy range $(1+z)E\sim (1+z)(E+\Delta E)$ may span multiple energy channels, we allocate the photon counts to each channel proportional to its width. The background spectra are shifted similarly to the source spectra they correspond to. 

Before reaching the X-ray instrument, the extragalactic light is first absorbed by the galactic medium. Such absorption, typically on the order of \qty{1e20}{cm^{-2}}, affects the soft X-ray band and causes suppression of intrinsic soft X-ray emission (which we are typically interested in). The correction of Galactic absorption cannot be applied on the PI spectrum directly, as the X-ray photon counts follow Poisson and cannot be scaled (see \cref{sec:basic_concepts}). Instead, we treat the Galactic absorption as an additional ``filter'' of the instrument. In practice, we take an energy-bin-wise product of the ARF and the Galactic absorption profile \tbabs~\citep[][]{Wilms+2000}, whose column density is specified by HEASARC's \nh~tool \footnote{\url{https://heasarc.gsfc.nasa.gov/cgi-bin/Tools/w3nh/w3nh.pl}}, given the RA and DEC of the target source. An example comparing the original ARF and Galactic absorption corrected ARF is shown in \cref{fig:ARF_plots}, from red solid to red dashed curve. We note that most of the sources ($\gtrsim90\%$, see \cref{fig:bii}) in our sample have relatively high Galactic latitude ($|l|>\ang{30;;}$), and therefore the column density given by \nh~should be secure.
For \ero, the energy range above $\sim \qty{0.2}{keV}$ is considered reliable \citep{Merloni+2024}, therefore, we truncate the ARF below $\qty{0.2}{keV}$. After these corrections on ARF, we multiply it to the RMF to obtain the Galactic-$N_\mathrm{H}$-corrected \textit{full response} $\mathbf{P}_{ij}$.

Since the PI spectrum and the \textit{full response} $\mathbf{P}_{ij}$ are coupled together through \cref{eq:basic}, a rest-frame correction must also be applied to $\mathbf{P}_{ij}$, to match the PI shifting. The $\mathbf{P}_{ij}$ matrix is nearly diagonal and two-dimensional, with rows corresponding to input model energy bins and columns to output detector channel energies. Each row sums to the effective area at the respective input energy, as encoded in the ARF's \texttt{SPECRESP} extension. The rest-frame correction of $\mathbf{P}_{ij}$ involves two steps:
\begin{enumerate}
    \item \textit{Horizontal shift} (along output channel energy axis): For each row, we shift the grids horizontally by $1+z$ while keeping the column bin edges fixed. When the shifted grid spans more than one channels (column bins), we allocate the grid value in each channel proportional to the overlapping width, following the same procedure as PI shifting. This step automatically broadens the energy dispersion (inverse of spectral resolution) by $1+z$.
    \item \textit{Vertical shift} (along input model energy axis): Then for each column, we shift the grids vertically by $1+z$, while keeping the row bin edges fixed. If a row bin overlaps with multiple shifted grids, a weighted average is computed based on the overlap width between the shifted grid and the row bin.
\end{enumerate}

\subsection{Spectral stacking} \label{sec:spec_stack}
To reach as high signal-to-noise ratio as possible, while preserving Poisson statistics, we sum the rest-frame PI spectra of all sources (indexed by $l$) directly, 
\begin{equation}
    C_i^\mathrm{stack}=\sum_l C_i^l\label{eq:stack_pi}
\end{equation}
The uncertainty of an energy bin with spectral counts $N$ follows a Poisson distribution. Unlike the routine of optical spectral stacking, we do not scale the spectra to a common flux level, which would deviate from Poisson statistic and pose additional challenges when estimating the uncertainty. 

Because the sum of Poisson processes with expectation $\lambda_i^l$ is a
Poisson process with the expectation $\sum_l \lambda_i^l$, we could sum up \cref{eq:basic} for all sources, and rewrite it as:
\begin{equation}
    C_i^\mathrm{stack}=\sum_lC_i^l=\mathrm{MODEL}+\mathrm{BKG}
\end{equation}
where
\begin{align}
    \mathrm{MODEL}&\sim\mathrm{Poisson}\left(\sum_l\sum_j\mathbf{P}_{ij}^l\times F_j^l\times\Delta t^l\times\Delta E_j\right)\\
    \mathrm{BKG}&\sim\mathrm{Poisson}\left(\sum_l b_i^l\right)\label{eq:stack1}
\end{align}
Here we have assumed that the responses have been mapped onto a common rest-frame energy grid.
We have not made the assumption $\sum_l R^l_{ij}\times A^l_{j} = \sum_l \left(R^l_{ij}\right)\times \sum_l \left(A^l_{j}\right)$, i.e., that the ARFs and RMFs can be stacked separately, as made in other spectral stacking tools.

We first treat the second term $\mathrm{BKG}$, i.e., the background contribution. A first-order approximation of $b_i^l$ is $B_i^l\times R_\mathrm{bkg}^l$, where $B_i^l$ is the background counts of the $l$th source, and $R_\mathrm{bkg}^l$ is the corresponding source-to-background ratio, calculated via $\frac{t_\mathrm{D}}{t_\mathrm{B}}\frac{b_\mathrm{D}}{b_\mathrm{B}}\frac{a_\mathrm{D}}{a_\mathrm{B}}$. Here $t$ is exposure time, $b$ is \verb|BACKSCAL| parameter (the expected size of internal background), $a$ is \verb|AREASCAL| parameter (the effective area scaling factor); $B$ for background extraction region, and $D$ for source extraction region\footnote{\url{https://heasarc.gsfc.nasa.gov/docs/xanadu/xspec/manual/node11.html}}. 

However, it is important to consider that $R_\mathrm{bkg}$ varies from source to source, meaning that $\sum_l B_i^l\times R_\mathrm{bkg}^l$ does not follow a Poisson process, and therefore we cannot simply estimate the background uncertainty as $\sqrt{\sum_l B_i^l\times R_\mathrm{bkg}^l}$.
We therefore adopt a different and robust method that nevertheless allows us to estimate the background uncertainty. We first divide the whole sample into 10 groups of similar $R_\mathrm{bkg}$. For each group, the background counts are summed, and the uncertainty is calculated assuming approximate Poisson statistics, with the error scaling as $\sqrt{N}$. Since the photon counts in each channel are sufficiently large now (valid for Gaussian approximation, $\mathrm{Poisson}(\lambda)\approx\mathcal{N}(\lambda,\lambda)$), we stack the 10 background spectra, scaling with the average source-to-background ratio ($\overline{R_\mathrm{bkg}^k}$) in each group, and calculate uncertainty in each channel with standard Gaussian error propagation. In summary, we approximate the background contribution as:
\begin{equation}
    \mathrm{BKG}\sim\mathcal{N}\left(\sum_l B_i^{l}\times R_\mathrm{bkg}^l,\ \sum_k(\overline{R_\mathrm{bkg}^k}^2\times\sum_{m_k}B_i^{m_k})\right)\label{eq:bkg}
\end{equation}
Note that we nevertheless estimate the expectation of background as $\sum_l B_i^l\times R_\mathrm{bkg}^l$, which is slightly more accurate than grouping into $k$ groups, given that $R_\mathrm{bkg}$ has some scatter even in the same group.

We then treat the first term of \cref{eq:stack1}, $\mathrm{MODEL}$, i.e., the source model contribution. The minimum assumption of spectral stacking is that \textit{all sources to be stacked share the same spectral shape}, and therefore we can rewrite $F_j^l$ as $\frac{F_j^l}{\overline{F_j}}\times\overline{F_j}$, where $\frac{F_j^l}{\overline{F_j}}$ is the normalization of source $l$ that does not vary across index $j$ (the input model energy). With that in mind, we further transform the first term of \cref{eq:stack1} into:
\begin{align}
    &\mathrm{Poisson}\left(\sum_l\sum_j\mathbf{P}_{ij}^l\times F_j^l\times\Delta t^l\times\Delta E_j\right)\\
    &=\mathrm{Poisson}\left(\sum_j(\sum_l\frac{F_j^l}{\overline{F_j}}\times\Delta t^l\times\mathbf{P}_{ij}^l)\times\overline{F_j}\times\Delta E_j\right)\\
    &=\mathrm{Poisson}\left(\sum_j(\sum_l W^l\times\mathbf{P}_{ij}^l)\times\overline{F_j}\times\Delta E_j\right)\\
    &=\mathrm{Poisson}\left(\sum_j\mathbf{P}_{ij}^\mathrm{stack}\times\overline{F_j}\times\Delta E_j\right)\label{eq:stack2}
\end{align}
with the response weighting factor $W^l$ being
\begin{equation}
    W^l=\frac{F_j^l}{\overline{F_j}}\times\Delta t^l
\end{equation}
In practice, we estimate $W^l$ from a pure data-driven way, based on \cref{eq:basic}:
\begin{equation}
    W^l=\frac{\sum_\mathrm{i=1.0\ keV}^\mathrm{2.3\ keV}(C_i^l-B_i^l\times R_\mathrm{bkg}^l)}{\sum_\mathrm{i=1.0\ keV}^\mathrm{2.3\ keV}\sum_j \mathbf{P}_{ij}^l\times\overline{F_j}\times\Delta E_j} \label{eq:weight}
\end{equation}
Ideally if the spectral counts is large enough and all sources really share the same spectral shape, $(C_i^l-B_i^l\times R_\mathrm{bkg}^l)/(\sum_j \mathbf{P}_{ij}^l\times\overline{F_j}\times\Delta E_j)$ should be nearly identical across the entire output channel energies ($i$). This is of course not the case in reality, and we therefore integrate over a broad energy range (rest-frame \qtyrange[range-phrase=--,range-units=single]{1.0}{2.3}{keV}) to obtain an average value. The choice of \qtyrange[range-phrase=--,range-units=single]{1.0}{2.3}{keV} is optimal for \ero, as it includes sufficient photon counts while deliberately exclude energies below $\qty{1}{keV}$ to minimize contamination by absorption. We assume a $\Gamma=2$ power-law with normalization of $1$ for $\overline{F_j}$. Choosing a different spectral shape (e.g., a power-law with $\Gamma=0$) changes the ultimate stacked response by $\lesssim 1\%$ and therefore does not affect our results. As we are only interested in the shape of the stacked spectrum, we choose the $\overline{F_j}$ normalization arbitrarily, and further normalize $W^l$ to $W^l_\mathrm{norm}=W^l/\sum_l W^l$. The final stacked full response, to be inserted back to \cref{eq:stack2}, is calculated as:
\begin{equation}
    \mathbf{P}_{ij}^\mathrm{stack}=\sum_l W^l_\mathrm{norm}\times \mathbf{P}_{ij}^l\label{eq:stack_rsp}
\end{equation}

We note the similarity between \cref{eq:stack2} and the first term of \cref{eq:basic}. This is to say, with the stacked PI spectrum, stacked background PI spectrum, and stacked response, we can proceed with spectral fitting just as we often do for the single source, to infer the real average model $\overline{F_j}$. Due to the Gaussian nature of the background, the PG-statistic (\texttt{pgstat}) is used.

\subsection{Implementation of code and notes}\label{sec:xstack_notes}
The complete spectral stacking procedure is implemented in our publicly available code, \xstack~(\url{https://github.com/AstroChensj/Xstack/}). To summarize, we stack the individual PI spectra as \cref{eq:stack_pi}, stack the individual background spectra and estimate uncertainty with \cref{eq:bkg}, and take the weighted sum of the full response following \cref{eq:weight,eq:stack_rsp}. All individual spectra and responses have already been shifted on the same rest-frame energy grid (\cref{sec:spec_shift}). The resulting stacked PI spectra, background PI spectra, and full responses (from which we can extract ARF and RMF) are fully prepared for spectral fitting. While the normalization of the stacked spectra is not physically meaningful due to the nature of our stack, our method robustly preserves spectral shapes, enabling reliable analysis of spectral trends and parameter dependencies.

Below we make some additional comments:
\begin{enumerate}
    \item The response weighting factor $W^l_\mathrm{norm}$ essentially characterizes the relative contribution of each source to the stacked spectrum. In \cref{fig:RSP_weight} we present the histogram of $W^l_\mathrm{norm}$ for the BH-mass and spec-z sample (see \cref{sec:smp_select} for definition of the two samples). The extremely-high-count sources (orange histogram) are also at the high end of the $W^l_\mathrm{norm}$ distribution, but have some dispersion due to the redshift distribution. The majority of sources sit around the median values, and no significant outliers are seen.
    \item By taking the inter-source flux variation into account, our data-driven response weighting method can characterize the mean sample spectrum with less bias than for example by exposure. Different approaches are tested with simulations in \cref{sec:simulation}. 
    \item From the stacked full response we can extract an ``effective'' ARF and RMF. The ARF can be derived by summing $\mathbf{P}_{ij}^\mathrm{stack}$ along the output channel energy axis, while dividing $\mathbf{P}_{ij}^\mathrm{stack}$ by the ARF yields the RMF. We plot the effective stacked ARF and RMF in \cref{fig:ARF_plots} (blue solid line) and \cref{fig:RMF_plots} (middle panel and blue curve in the right panel) respectively, in comparison with the ARF and RMF for a single source. 

    The shape of the stacked ARF is clearly different from that of a single ARF. For the stacked RMF, we can also see two major differences from the single RMF: 1) the off-axis features, prominent in the single RMF, are now blurred and smoothed in the stacked RMF; 2) the diagonal of the stacked RMF becomes thicker than that in the single RMF. Indeed, by fitting each row (input model energy) of the RMF matrix with a Gaussian, the right panel of \cref{fig:RMF_plots} shows how the spectral energy dispersion (characterized by FWHM) in the stacked RMF are systematically larger than those in the single RMF. This apparent degradation of the spectral resolution mainly stems from the response shifting process (rather than the stacking process), where the rest-frame energy dispersion $\mathrm{FWHM}^\mathrm{rest}$ at $E_{j}$ is actually $1+z$ times the observer-frame energy dispersion $\mathrm{FWHM}^\mathrm{obs}$ at $E_j/(1+z)$, and from the relation between dispersion and energy for a single RMF (red curve in right panel of \cref{fig:RMF_plots}) it can be read that $\mathrm{FWHM}^\mathrm{obs}_{E_{j}/(1+z)}>\mathrm{FWHM}^\mathrm{obs}_{E_{j}}/(1+z)$. Therefore, $\mathrm{FWHM}^\mathrm{rest}_{E_{j}}=(1+z)\cdot\mathrm{FWHM}^\mathrm{obs}_{E_{j}/(1+z)}>\mathrm{FWHM}^\mathrm{obs}_{E_{j}}$.
\end{enumerate}

\section{Robustness checks}\label{sec:robust_stack}
In this section, we test the robustness of our results. We first perform single-power-law simulation to validate our X-ray spectral stacking procedure in \cref{sec:simulation}. We check whether our choice of K-correction slope would result our results in \cref{sec:kcorr}. We also check whether several sample selection bias would affect our conclusions in \cref{sec:selection_bias}. Finally, we compare the brightest 5 sources with the stacked spectra in each of the $3\times 3$ $M_\mathrm{BH}-\lambda_\mathrm{Edd}$ bin, as a sanity check for our optical-UV-Xray stacking.

\subsection{Simulation to validate stacking}\label{sec:simulation}
\begin{figure}
    \centering
    \includegraphics[width=0.9\linewidth]{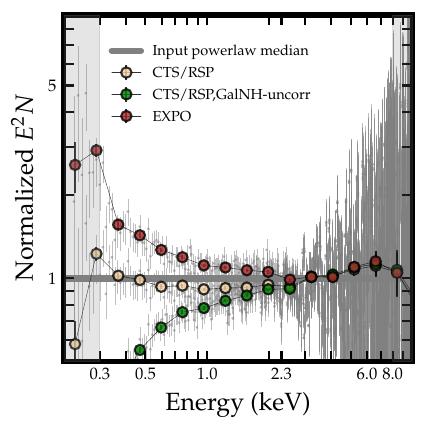}
    \caption{Simulation results. For a family of mock power-laws with varying photon indices (centered at 2), redshift, Galactic absorption, exposure time, background, and response (all taken from real values), \xstack~produces the stacked spectrum (assuming our data-driven response weighting factors) as yellow dots (``CTS/RSP''), resembling the input median power-law well. For comparison, we present the stacked spectrum without Galactic absorption correction (green dots, ``CTS/RSP,GalNH-uncorr''), and the stacked spectrum taking exposure as response weighting factors (red dots, ``EXPO''). They both reproduce much more biased spectral shapes.}
    \label{fig:simulation}
\end{figure}

To validate our stacking procedure, we perform simulation using \xspec~command \fakeit. For simplicity, we adopt a power-law absorbed by the Galactic medium, \verb|TBabs*zpowerlw|, as the input model. The photon index is assumed to follow a Gaussian distribution with a mean of 2 and a standard deviation of \num{0.2}. The redshift, Galactic absorption column density, 
\qtyrange[range-phrase=--,range-units=single]{0.3}{2.3}{keV} observer-frame flux, exposure time, background and ARF, are based on real sources, sampled from \num{500} random sources in our dataset. These parameters ensure that the simulated Galactic-absorbed power-laws share similar spectral qualities to the real sample. The median of input spectral shape (before being absorbed by Galactic medium) is plotted in \cref{fig:simulation} as gray horizontal ($\Gamma=2$) solid line. We then apply three different methods to stack them, and present the stacked spectra in \cref{fig:simulation}. Common to the three methods, we sum the rest-frame PI counts directly. For the yellow points (labeled ``CTS/RSP'') we apply our data-driven response weighting method (\cref{eq:weight}) and Galactic-absorption correction, whereas for the green points (labeled ``CTS/RSP, GalNH-uncorr'') we use the same response weight, but do not apply Galactic absorption correction. For the red points (labeled ``EXPO''), we weigh each response by exposure time. 

The stacked spectrum generated using our data-driven response weights (yellow curve, $\Gamma\sim 1.96$) closely aligns with the input median (gray horizontal line, $\Gamma=2$), highlighting the reliability of our stacking method. Below \qty{0.3}{keV}, we observe minor artifacts that likely originate from low source completeness in this energy range. And this is also the reason why we have discarded below \qty{0.3}{keV} in the scientific analysis (\cref{sec:spec_fit}). 

In the meantime, we see that the two other methods all produce significantly biased spectral shape compared to input, especially towards the soft end. Weighting response by exposure time is a rather simplified method, and would only be identical to our data-driven method when all sources have the same flux in the \qtyrange[range-phrase=--,range-units=single]{1.0}{2.3}{keV} band. As can be seen, while this approach reproduces the hard X-ray part reasonably well, it produces a large artifactual excess in the soft band, resulting in an overall softer ($\Gamma\sim\num{2.12}$) spectrum than expected ($\Gamma=2$). This emphasizes the importance of accounting for inter-source flux variation during stacking. On the other side, from the green points we see that the Galactic absorption significantly suppress the stacked spectrum up to $\sim\qty{3}{keV}$, and thus highlighting the need to correct for it during the stacking.

\subsection{Validity of constant SED slope for UV K-correction}\label{sec:kcorr}
\begin{figure}
    \centering
    \includegraphics[width=1.0\linewidth]{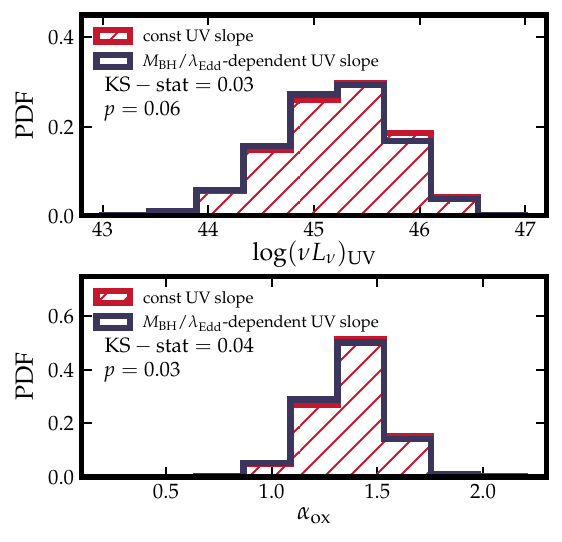}
    \caption{\textit{Upper panel}: the comparison between the UV luminosity derived with K-correction assuming a constant UV SED slope $\alpha=-0.65$, and that assuming a $M_\mathrm{BH}$-$\lambda_\mathrm{Edd}$-dependent slope. The KS test statistics as well as p-values are also listed. We generally do not see significant differences between the two distributions. \textit{Lower panel}: same as upper panel, but comparing $\alpha_\mathrm{ox}$ instead.}
    \label{fig:modified_UV}
\end{figure}
To derive the rest-frame UV luminosity of each source, we apply a K-correction assuming a constant UV SED slope of $\alpha=-0.65$. However, as shown by \citealt{Hagen+2024} and also our \cref{fig:agnsed_model}, the UV slope actually evolves with $\lambda_\mathrm{Edd}$ (and slightly with $M_\mathrm{BH}$). Ideally, one could iteratively stack the UV SED within bins of $M_\mathrm{BH}$ and $\lambda_\mathrm{Edd}$, updating the assumed SED slope for K-correction at each step, until convergence to an ``optimal'' value is achieved. This would however require $\lambda_\mathrm{Edd}$, $M_\mathrm{BH}$, and \g, \u~photometry from \sdss, which is not applicable to all sources in our sample (especially the ``spec-z'' sample). We therefore simply adopt a uniform constant slope for the K-correction across all sources, regardless of their physical properties. Below, we perform simple tests to validate that our conclusions are robust against this choice.

We select a representative subsample from our ``BH-mass'' and ``spec-z'' sample, consisting of sources located within the $3\times3$ grid in \cref{fig:MBHEdd_dist}. Since we have stacked UV spectra for this subsample (\cref{fig:agnsed_model}), we can use them to update our knowledge of the $M_\mathrm{BH}$-$\lambda_\mathrm{Edd}$-dependent UV SED slopes. (For simplicity, we do not perform further iterations to refine the slopes.) Using these updated slopes, we recalculate the K-corrected UV luminosity $\log(\nu L_\nu)_\mathrm{UV}$ and $\alpha_\mathrm{ox}$ for each source, and compare them to the values obtained under the assumption of a constant slope $\alpha=-0.65$. The comparisons are shown in \cref{fig:modified_UV}, along with KS test statistics and p-values. From the upper panel we see that the distribution of $\log(\nu L_\nu)_\mathrm{UV}$ remains nearly unchanged even after applying the updated slopes, with only marginal differences ($p=0.06$). A similar conclusion holds for $\alpha_\mathrm{ox}$, as shown in the lower panel. Therefore our conclusions in \cref{sec:pheno_data,sec:pheno_fit_results} are not affected. 

Finally, we examine whether our conclusions based on \agnsed~fit (\cref{sec:agnsed_fit_results}) are affected by applying the updated, $M_\mathrm{BH}$-$\lambda_\mathrm{Edd}$-dependent UV slopes for K-correction. We find that the changes in the best-fit \agnsed~parameters (notably, the warm corona radius $R_\mathrm{WC}$) are all within $1\sigma$ uncertainties. This is in fact understandable, as for our UV stacking pipeline (\cref{fig:agnsed_model}), we have already assigned additional uncertainties to the stacked UV SED to account for potential systematics, including the uncertainties in K-correction slope.

\subsection{Robustness against sample selection bias}\label{sec:selection_bias}
We also check the robustness of our conclusions against potential sample selection bias. We first account for the \nuv~non-detected sources in each of the $3\times3$ grid in \cref{fig:agnsed_model}, that is, we select all \sdss-\ero~sources within the \galex~field of view, regardless of \nuv~detected or not. In stacking optical-UV-Xray SED, these \nuv~non-detected sources contribute \ero~X-ray, \sdss~\g~and \u~data, but no \galex~\nuv~flux, leading to a downward trend in the UV. After fitting the resulting SED with the same procedure as \cref{sec:fit_agnsed}, we find very small changes in all \agnsed~parameters, which is expected as the overall \nuv~completeness in our BH-mass sample is as high as $\sim 93\%$ (\cref{sec:smp_select}). Secondly, we check if our conclusions still hold for a lower redshift subsample. Although the majority of sources in our BH-mass sample have relatively low redshift ($z<1$), there is still a portion with high redshift ($z>1$), reaching the EUV regime and thereby bringing in more IGM absorption, potentially biasing the entire SED. Focusing on a subsample of $z<1$ and doing the spectral fitting again on the $3\times3$ grid, we find that correlations between warm/hot corona radius and $\lambda_\mathrm{Edd}$ and $M_\mathrm{BH}$ are still preserved.

\subsection{Brightest 5 sources in each $3\times 3$ grid}\label{sec:single_source}
\begin{figure*}
    \centering
    \includegraphics[width=0.8\linewidth]{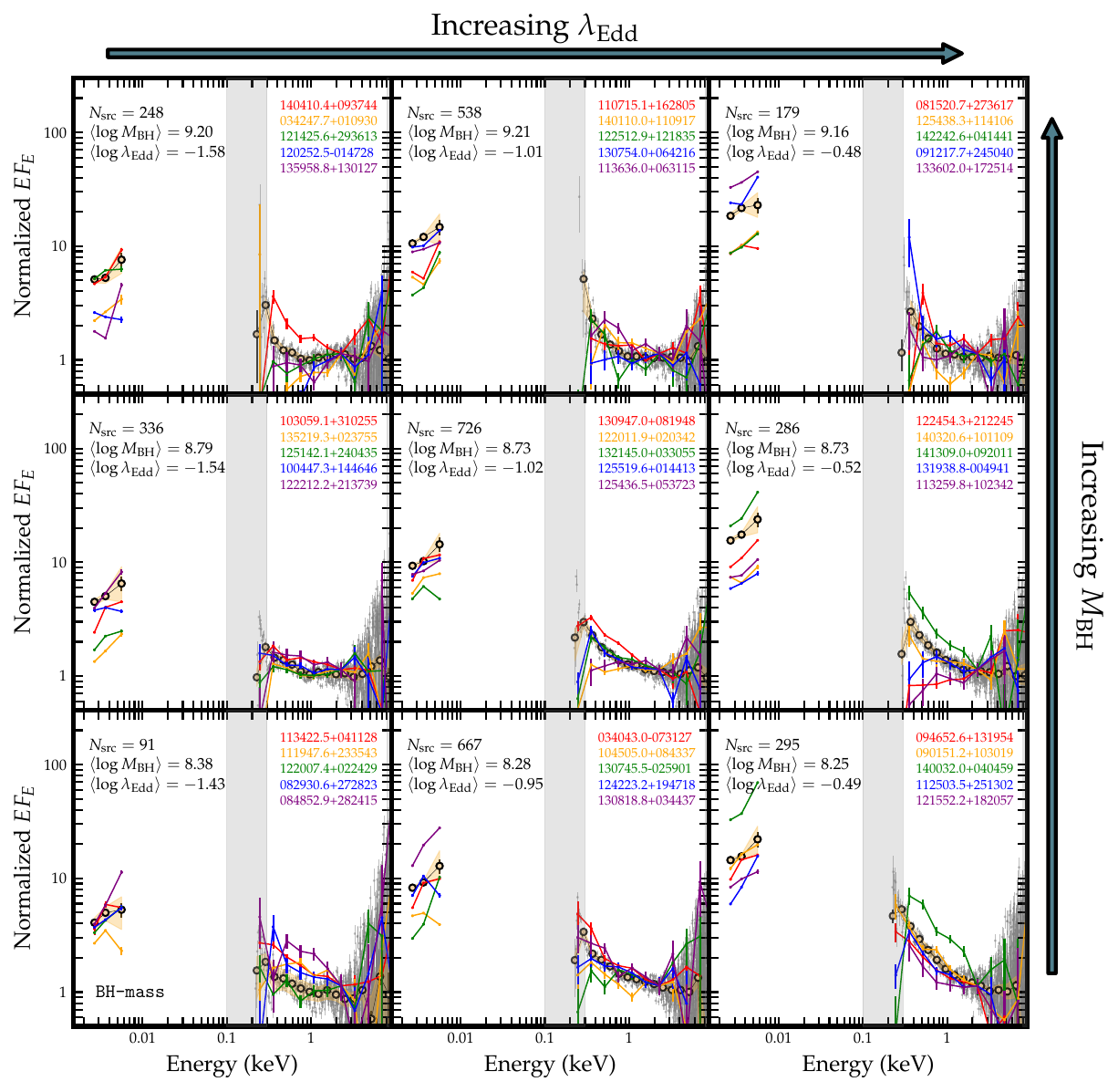}
    \caption{Like \cref{fig:agnsed_model}, but also showing the individual spectra of the brightest 5 sources in each $M_\mathrm{BH}$ -- $\lambda_\mathrm{Edd}$ bin as colored curves. These generally show the same trends as the stacked re-binned data (yellow dots with error bars) and unbinned data (gray error bars).}
    \label{fig:agnsed_single}
\end{figure*}
We compare the spectra of the brightest 5 sources to the stacked SED in each of the $3\times3$ grid in \cref{fig:agnsed_single}. The individual sources are shifted to rest-frame, in the same way as the stacked spectra illustrated in \cref{sec:spec_shift} (X-ray) and \cref{sec:UV_stack} (UV). Since the individual spectra have rather poor spectral quality at $4\ \mathrm{keV}$, we normalize them at $\sim 2\ \mathrm{keV}$ to the same flux of the normalized stacked spectrum. Although these bright sources still have very limited photon counts ($\sim 200$ within \qtyrange[range-phrase=--,range-units=single]{0.2}{10.0}{keV}) for reliable spectral modeling, it is noticeable that these individual sources generally follow the same trend as those observed in the stacked spectra. In particular, the soft excess becomes stronger for increasing $\lambda_\mathrm{Edd}$ and decreasing $M_\mathrm{BH}$, and the UV-to-Xray ratio becomes stronger for increasing $\lambda_\mathrm{Edd}$ but does not correlate well with $M_\mathrm{BH}$.

\section{Star formation contamination in low $\lambda_\mathrm{Edd}$ -- high $M_\mathrm{BH}$ bins}\label{sec:sf_contamination}
\begin{figure}
    \centering
    \includegraphics[width=0.9\linewidth]{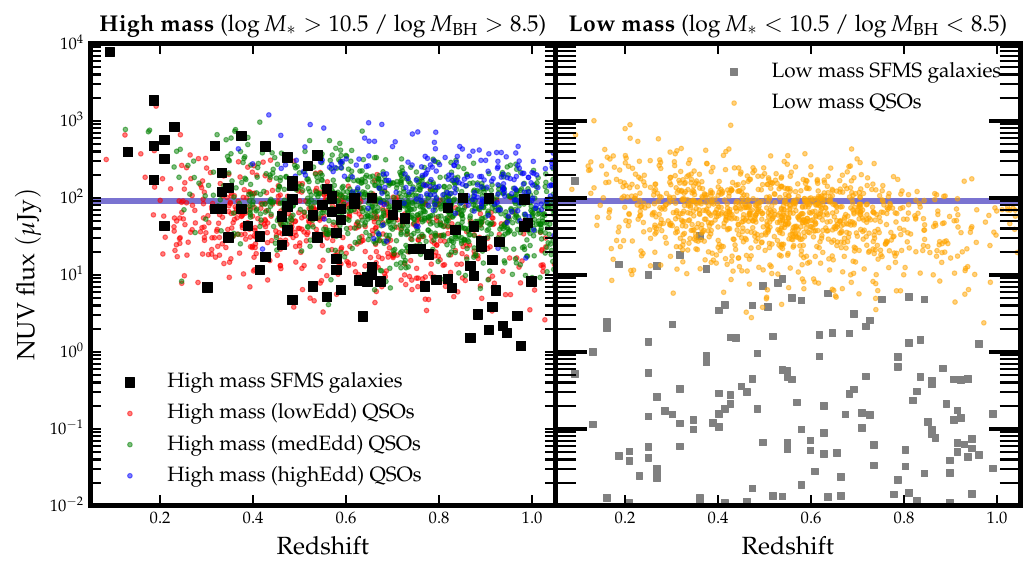}
    \caption{Estimated NUV flux of SFMS galaxies with high mass (black points in the left) and low mass (gray points in the right), and the observed NUV flux of quasars in our sample with high mass (red for low $\lambda_\mathrm{Edd}$, green for medium $\lambda_\mathrm{Edd}$, blue for high $\lambda_\mathrm{Edd}$) and low mass (yellow).}
    \label{fig:SFMS_QSO_Nflux}
\end{figure}

In \cref{fig:SFMS_QSO_Nflux} we present the estimated NUV flux (in \unit{\mu Jy}) of galaxies on the star formation main sequence (SFMS), shown as black and gray points. These points are derived from an ensemble of GRAHSP \citep{Buchner+2024} models, selected with star formation rate \verb|SFR|$<\qty{300}{M_\odot.\mathrm{yr}^{-1}}$, specific star formation rate $-10.5<$\verb|log_sSFR|$<-9.5$, and dust attenuation \verb|E(B-V)|$<0.03$ to approximate the SFMS. Additionally, an AGN luminosity cut \verb|log_L_AGN|$<40$ is applied to exclude AGN contributions. 

Assuming stellar-to-SMBH mass ratio of $\sim 200$, we plot in the left panel the predictions for high mass galaxies ($\log M_*>10.5$) from GRAHSP, alongside corresponding quasars (AGN+host) with $\log M_\mathrm{BH}>8.5$ from our sample. We find that the NUV flux from SFMS massive galaxies can reach up to \qty{200}{\mu Jy}, which is similar to the low $\lambda_\mathrm{Edd}$ quasars (red) in our sample, and only slightly below the medium (green) and high (blue) $\lambda_\mathrm{Edd}$ quasars. On the contrary, the right panel shows the fluxes from low mass SFMS galaxies ($\log M_*<10.5$) and corresponding quasars with $\log M_\mathrm{BH}<8.5$ from our sample (yellow, with no further differentiation by $\lambda_\mathrm{Edd}$). Here, the fluxes from SFMS galaxies are consistently smaller than those from the quasars. These suggest that the real AGN contributions in the low $\lambda_\mathrm{Edd}$ -- high $M_\mathrm{BH}$ could be significantly smaller, if their host galaxies really lie on SFMS.



\end{document}